\documentclass[twocolumn, times, tighten, appendixfloats]{aastex7}

\usepackage{hyperref}
\usepackage{romannum}
\usepackage{longtable}

\begin{document}

\pagenumbering{arabic}

\title{On The Very Bright Dropouts Selected Using the James Webb Space Telescope NIRCam Instrument}

\author[0000-0001-7957-6202]{Bangzheng Sun} 
\affiliation{Department of Physics and Astronomy, University of Missouri - Columbia \\
701 S College Avenue \\
Columbia, MO 65201, USA}
\email{bangzheng.sun@mail.missouri.edu}

\author[0000-0001-7592-7714]{Haojing Yan}
\affiliation{Department of Physics and Astronomy, University of Missouri - Columbia \\
701 S College Avenue \\
Columbia, MO 65201, USA}
\email{yanha@missouri.edu}

\begin{abstract}

The selection of candidate high-redshift galaxies using the dropout technique
targeting the Lyman-break signature sometimes yields 
very bright objects that are 
too luminous to be easily explained if they are indeed at the 
expected redshifts. Here we present a systematic study of very bright dropouts 
selected through successive bands of the NIRCam instrument onboard the James Webb 
Space Telescope (JWST). Using the public NIRCam data in four blank fields over
500~arcmin$^2$, 300 such objects were found. They have F356W magnitudes 
$<25.1$~mag or $<26.0$~mag depending on the dropout passband, and the 
majority of them ($>80\%$) have very red F115W$-$F356W colors $> 2.0$~mag, 
qualifying them as ``extremely red objects'' (EROs). We focus on 
137 objects that also have mid-IR observations from the JWST MIRI instrument. 
Their spectral energy distribution analysis shows that 
these objects 
are dominated by low-redshift ($z\sim1$--4) galaxies ($\gtrsim67\%$). 
However, a non-negligible fraction ($\gtrsim7\%$) could be at high redshifts. 
Seven of our objects have secure spectroscopic redshifts from JWST NIRSpec 
identifications, and the results confirm this picture: while six are 
low-redshift galaxies ($z\approx3$), one is a known galaxy at $z=8.679$ 
{(with $M_{\rm UV}=-22.4$~mag and stellar mass $M_*=10^{9.1}M_\odot$)}
recovered in our sample. 
{
In light of recent theoretical models on early galaxy 
formation, this confirmed high-redshift galaxy does not pose a challenge. 
However, as our sample contains very luminous high-redshift candidates in the regime 
still underexplored 
($M_{\rm UV}\leq -23$~mag and $M_*>10^{10.5}M_\odot$), 
spectroscopic identifications are necessary to ensure 
they do not create tension with these new models. 
}

\end{abstract}


\section{Introduction} \label{sec:intro}

   Various kinds of ``extremely red objects'' (EROs) have been selected over 
the past three decades by using different color indices involving different
infrared (IR) bands as the technologies evolve. The early discoveries were
mostly made by comparing the ground-based observations in $K$-band 
($\sim$2.2~$\mu$m) and in an optical band
\citep[e.g.,][]{Elston1988, HuRidgway1994, Thompson1999, Scodeggio2000},
although occasionally deep data in the less red F160W band
($\sim$1.6~$\mu$m) of the Hubble Space Telescope (HST) Near Infrared Camera 
and Multi-Object Spectrometer (NICMOS) were also used 
\citep[e.g.,][]{YanL2000}. The general interpretation of such EROs is that 
they are predominantly early-type galaxies at $z\approx 1$ and that their red 
IR-to-optical colors are due to their old stellar populations. After the 
launch of the Spitzer Space Telescope, the ERO selection proceeded to using 
the Infrared Array Camera (IRAC) 3.6 and/or 4.5~$\mu$m channels in IR and the 
HST Advanced Camera for Surveys (ACS) bands in optical
\citep[e.g.][]{Yan2004c}.
Such EROs are also mostly evolved galaxies but are at higher redshifts of
$z\approx 2$--4. The extension in the IR wavelengths by IRAC also allowed the 
ERO selection through IR colors such as $K_s - 4.5\mu$m involving ground-
based $K_s$ band \citep[e.g.,][]{WangWH2012} and ${\rm F160W} - 4.5\mu$m 
involving the reddest band (F160W) of the HST Wide Field Camera 3 (WFC3)
\citep[e.g.,][]{Caputi2014, WangTao2016, AP2019}.
The latter ones are also referred to as the ``HST-dark'' objects because they
are extremely weak or even invisible in the deep images taken in any HST 
bands. Interestingly, the EROs selected using IR color indices consist of 
not only early-type galaxies dominated by passively evolving stellar
populations but also extremely dust-reddened galaxies with strong ongoing star
formation embedded by dust \citep[e.g.,][]{WangTao2019}.
In fact, dusty starbursts invisible at $\lambda < 2$~$\mu$m
were already known to exist, for instance, the historical submillimeter galaxy
(SMG) HDF 850.1 \citep{Cowie2009, Walter2012}. Many similar SMGs were later
revealed \citep[e.g.,][]{Zhou2020, GG2022, Xiao2023}, making them a distinct 
population among EROs at $z>4$ awaiting further exploration.

   In a broad sense, the selection of high-redshift (high-$z$) galaxies also
relies on identifying red colors, which are due to the Lyman-break signature 
in their spectral energy distributions (SEDs). This is caused by the discrete
neutral hydrogen clouds along the sightline that absorb the photons bluer
than the Lyman-limit (rest frame $\lambda<912$\AA) and those coinciding with the
Ly$\alpha$ line wavelength (rest frame $\lambda=1216$\AA).
These absorptions create a sharp discontinuity (``Lyman break'') 
in the spectrum of a
high-redshift (high-$z$) galaxy, resulting in a red color index in two 
adjacent bands that straddle the Lyman break. For this reason, a high-$z$ 
galaxy would appear to ``drop out'' from the band to the blue side of the 
break and 
{be visible in bands redder than the break, }
and this is the basis of the classic 
``dropout'' selection of galaxies at $z\gtrsim 3$
\citep[][]{Steidel1992, Steidel1993, Steidel1995}.
At $z>5$, the cumulative Ly$\alpha$ line absorptions
(``Ly$\alpha$ forest'') are so strong that the Lyman break occurs at the rest 
frame 1216\AA. In practice, candidate high-$z$ galaxies selected based on the 
Lyman-break signature always have contaminants from red galaxies at low 
redshifts (``low-$z$ interlopers''), 
and the most extreme ones could be the aforementioned EROs
{\citep[see, e.g., ][]{AH23,Zavala23}.}

    JWST offers unprecedented sensitivity and spatial resolution in the IR
wavelengths. It has pushed the redshift record of galaxies to $z=14.32$
\citep[][]{Carniani2024}, and thousands of candidates have been selected at 
$z\gtrsim 7$ using its NIRCam instrument.
In the meantime, it has also brought the study of EROs to a new level 
\citep[e.g.,][]{Rodighiero2023, Nelson2023a, GG2023, Barrufet2023, 
Gibson2024}.
When the hunt for high-$z$ galaxies moves forward to higher and higher
redshifts, the selected candidates bear more and more similarity to EROs. As 
an example, galaxies at $z\gtrsim 12$ should be invisible at
$\lambda < 1.5$~$\mu$m, which could be regarded as being HST-dark.

   On the face value, all known EROs are IR-bright. For instance, typical 
HST-dark galaxies have AB magnitudes of $\leq 24$~mag at 3--5~$\mu$m. 
Therefore, one might think that the IR brightness could be used to 
distinguish the two populations. However, EROs being IR-bright reflects more 
the limitation of the technology at the time when they were first studied 
than their intrinsic properties. In this context, it is important to study how 
these two populations overlap in the new JWST era, starting from the bright 
end. In this work, we aim at the very bright high-$z$ candidates selected 
using the NIRCam data following the classic dropout technique and investigate 
how they could be affected by low-$z$ EROs. In order to make our analysis  
robust, we focus on the fields where the mid-IR imaging data from the JWST 
MIRI instrument are also available. 

    The structure of our paper is as follows. We describe the relevant NIRCam
and MIRI data in Section 2 and the photometry in Section 3, respectively. The 
selection of bright dropouts in the NIRCam bands is detailed in Section 4, 
followed by the analysis of their SEDs in Section 5. Based on their 
photometric redshifts ($z_{\rm phot}$), these bright dropouts are broadly 
categorized in Section 6 as being potential high-$z$ galaxies and possible 
low-$z$ EROs. A small fraction of our objects have spectroscopic 
confirmations, which we show in Section 7. We discuss our results in 
Section 8 and conclude with a summary in Section 9. All magnitudes quoted in 
this work are in the AB system \citep[][]{OG1983_abmag}. 
All coordinates are in the ICRS frame and Equinox 2000. 
We adopt a flat $\Lambda$CDM cosmology with $H_0=71$ km s$^{-1}$ Mpc$^{-1}$, $\Omega_m=0.27$, and $\Omega_\Lambda=0.73$.

\section{JWST NIRCam and MIRI Imaging Data}

  We utilize four JWST fields that have both NIRCam and MIRI data,
which are summarized in Table~\ref{tab:allfields}. These are from the Public 
Release IMaging for Extragalactic Research (PRIMER) program (PI J. Dunlop) 
in the COSMOS and the UDS fields, the Cosmic Evolution 
Early Release Science Survey \citep[CEERS;][]{Finkelstein2025_ceers} in the EGS 
field, and the JWST Advanced Deep Extragalactic Survey \citep[JADES;][]
{Eisenstein2023} in the GOODS-S region. 
These data and their reduction are briefly described below.

\begin{table*}[]
    \centering
    \tiny
    \begin{tabular}{ccccccccccc}
        Field & R.A. (deg) & Decl. (deg) & Instrument & Pipeline & Context & Area (arcmin$^2$) & MIRI Overlap (arcmin$^2$) & Exposure (ks) \\ \hline
        COSMOS & 150.12299 & 2.34764 & NIRCam & 1.10.2 & 1089 & 137.13 & 67.80 & $\sim$2.5 \\
         & & & MIRI & 1.12.5 & 1183 & 100.58 & - & $\sim$1.7 \\ 
         \hline
        UDS & 34.35004 & $-$5.20001 & NIRCam & 1.10.2/1.11.3 & 1089/1106 & 185.83 & 76.91 & $\sim$1-4 \\ 
         & & & MIRI & 1.12.5 & 1180 & 112.51 & - & $\sim$1.7 \\ 
         \hline
        CEERS & 215.00545 & 52.93451 & NIRCam & 1.9.4 & 1046 & 86.45 & 3.72$-$7.69 & $\sim$2.5 \\ 
         & & & MIRI & 1.12.5 & 1183 & $\sim5-16$ & - & $\sim$1-8 \\ 
         \hline
        JADES (deep) & 53.16444 & $-$27.78256 & NIRCam & 1.12.5 & 1180 & 25.50 & 20.83 & $\sim$14-60 \\ 
        JADES (medium) & 53.07011 & $-$27.90011 & NIRCam & 1.12.4 & 1140 & 37.94 & 15.99 & $\sim$1.5-40 \\ 
        GOODS-S & 53.14027 & $-$27.82140 & MIRI & 1.13.4 & 1188 & 32.90 & - & $\sim$0.6-142 \\ 
        \hline
    \end{tabular}
    \caption{Summary of the fields used in this work, which have data from 
    both NIRCam and MIRI. The nominal exposure times in these fields are 
    given under ``Exposure''. The equatorial coordinates are at the field 
    centers, and the sizes of the overlapped regions between the NIRCam and
    the MIRI coverages are indicated in the 
    {\bf ``MIRI Overlap'' column. }
    The version of the JWST data reduction pipeline that we used and 
    the relevant context file (``pmap'') are listed under ``Pipeline'' and 
    ``Context'', respectively. The astrometry that we adopted in the COSMOS, 
    UDS and CEERS fields is that from the HST CANDELS survey, while that in 
    the JADES fields are based on the GAIA DR3.}
    \label{tab:allfields}
\end{table*}

\begin{table*}[]
    \centering
    \footnotesize
    \begin{tabular}{cccccccccc}
        Field (NIRCam) & F090W & F115W & F150W & F200W & F277W & F335M & F356W & F410M & F444W \\ \hline
        CEERS & - & 28.5 & 28.5 & 29.4 & 29.5 & - & 29.0 & 28.8 & 29.3 \\ 
        UDS & 28.1 & 28.1 & 28.3 & 28.4 & 29.0 & - & 29.2 & 28.4 & 28.9 \\ 
        COSMOS & 28.2 & 28.2 & 28.4 & 28.6 & 29.2 & - & 29.3 & 28.6 & 29.0 \\ 
        JADES (deep) & 30.4 & 30.7 & 30.6 & 30.6 & 31.0 & 30.4 & 30.8 & 30.5 & 30.8 \\ 
        JADES (medium) & 29.2 & 28.5 & 29.0 & 28.8 & 29.2 & 30.0 & 29.8 & 29.6 & 29.0 \\ 
        \hline

    \\ 
        Field (MIRI) & F560W & F770W & F1000W & F1280W & F1500W & F1800W & F2100W & F2550W \\ \hline
        CEERS & 26.8 & 26.0 & 26.5 & 26.1 & 25.8 & 25.0 & 25.4 & - \\ 
        UDS & - & 27.3 & - & - & - & 25.3 & - & - \\ 
        COSMOS & - & 27.3 & - & - & - & 25.3 & - & - \\ 
        GOODS-S & 27.1 & 27.5 & 26.3 & 26.1 & 26.1 & 25.0 & 25.1 & 23.2 \\ 
        \hline
    \end{tabular}
    \caption{
    {Averaged 2~$\sigma$ depth in each passband for the fields in 
    Table~\ref{tab:allfields}, measured in a $r=0$\farcs2 circular aperture on 
    the RMS maps. The limits adopted for a specific object in case of
    non-detection are based on the local depths at the source position, which
    could be different from these averaged depths.  
    }
    }
    \label{tab:field_depths}
\end{table*}

\subsection{Data Description}

$\bullet$ {\bf PRIMER in COSMOS and UDS}: two shallow but wide fields. The 
total areas covered by NIRCam are 137.13 and 185.83~arcmin$^2$, respectively, 
and those covered by MIRI are 100.58 and 112.51~arcmin$^2$, respectively. 
The PRIMER program executed the MIRI observations as the primary and the 
NIRCam ones as the coordinated parallel while maximizing the overlapping 
areas between the two. In the end, the overlapping areas between the NIRCam 
and MIRI observations are 67.80 and 76.91~arcmin$^2$ for these two fields, 
respectively.
The NIRCam observations utilized eight passbands: F090W, F115W, F150W, and 
F200W in the short-wavelength channel (SW), and F277W, F356W, F410M, and F444W
in the long-wavelength channel (LW). The MIRI observations were in F770W and 
F1800W. In these two fields, we selected dropouts in F090W, F115W, F150W, 
F200W, and F277W. To assist the F090W dropout selection, we integrated the 
HST ACS data from the CANDELS survey \citep[][]{Grogin2011,Koekemoer2011} in 
both COSMOS and UDS. 

$\bullet$ {\bf CEERS in EGS}: another shallow but wide field.
It was observed in seven NIRCam bands: F115W, F150W, and F200W in the SW 
channel, and F277W, F356W, F410M, and F444W in the LW channel. The NIRCam
coverage is 86.45~arcmin$^2$.
The MIRI observations were made in smaller areas and were in seven 
bands: F560W, F770W, F1000W, F1280W, F1500W, F1800W, and F2100W. 
However, the MIRI footprints are non-overlapping for all these bands,
and different parts of the field are covered only by certain sets
of filters: (1) F560W and F770W (7.69~arcmin$^2$ overlapping with NIRCam), 
(2) F1000W, F1280W, F1500W, and F1800W (3.72~arcmin$^2$ overlaping with 
NIRCam), (3) F1500W only (5.15~arcmin$^2$ overlapping with NIRCam), and 
(4) F2100W only (no overlap with NIRCam). 
Due to the lack of the F090W data, we only searched for the F115W, F150W, 
F200W, and F277W dropouts. We utilized the ancillary HST ACS data from the 
CANDELS survey to assist the F115W dropout selection. 

$\bullet$ {\bf GOODS-S}: the main part of the data are from the JWST Advanced 
Deep Extragalactic Survey \citep[JADES;][]{Eisenstein2023}, which made the 
NIRCam observations in two areas within the GOODS-S field at two different 
depths, ``deep'' and ``medium'' (hereafter JADES-deep and JADES-medium). The 
data were taken in 9 NIRCam bands: F090W, F115W, F150W, and F200W in the SW 
channel, and F277W, F335M, F356W, F410M, and F444W in the LW channel. These 
data are all much deeper than PRIMER and CEERS. We also integrated the NIRCam 
data from the Pure Parallel Wide Area Legacy Imaging Survey 
\citep[PANORAMIC, PID 2514;][]{Williams2024} and those from PID 3215 {(PI D. Eisenstein)} 
into the JADES-medium field. 
The final NIRCam images cover 25.50 and 37.94~arcmin$^2$ in JADES-deep and
JADES-medium, respectively. For MIRI, we combined all the public observations 
overlapping the JADES coverage into one mosaic per band, which include those 
from PID 
1207 {(PI G. Rieke)}, 
1283 {(PI H. Norgaard-Nielsen)}, 
and 1180 {(PI D. Eisenstein)}. Eight MIRI bands were utilized by these programs: 
F560W, F770W, F1000W, F1280W, F1500W, F1800W, F2100W, and F2550W, and the final
MIRI images cover 32.90~arcmin$^2$ in all eight bands. 
The overlapping areas between NIRCam and MIRI are 20.83 and 15.99~arcmin$^2$ in
JADES-deep and JADES-medium, respectively.
{Unfortunately, another JADES field, GOODS-N, did not have overlapped 
NIRCam and MIRI imaging data at the time of this study; 
therefore, it was not included in this work. }

\subsection{Data Reduction}

    We reduced all the data in this study on our own using the JWST pipeline
\citep{Bushouse24_jwppl}. The NIRCam data reduction followed the procedures 
outlined in \citet[][]{Yan_highz_23} and \citet[][]{YSL2024}. 
For MIRI, we ran through the similar 
procedures but with four changes: (1) in the {\tt calwebb\_detector1} step, we
set {\tt jump.find\_showers} to ``True'' to remove the large residuals on 
single MIRI exposures due to strong cosmic ray events; (2) for the products 
after the stage 2 pipeline process, we followed the recipe by \citet[][]{Yang2023} 
to remove the stripe-like noise pattern in horizontal and/or vertical 
directions; 
{(3) following the same recipe, we then constructed a background image on a per-observation and per-filter basis 
by first masking the source and then taking the median of all single exposures, 
which was used as the template to remove the remaining noise patterns; }
and (4) we excluded the 
coronagraph areas in each individual exposure by masking their pixels to
{\tt DO\_NOT\_USE} in the data quality array.



\section{Photometry}

    All the final NIRCam and MIRI mosaics have a pixel scale of 0\farcs06, 
which translates to the AB magnitude zero-point of 26.581. We used 
\texttt{SExtractor} \citep{Bertin1996} for source extraction and 
photometry. We treated the NIRCam and the MIRI images separately because the
latter have much larger point spread functions (PSFs).

\subsection{NIRCam photometry for dropout selection}

    The dropout selection involves the NIRCam data but not the MIRI data. For 
each field, we ran \texttt{SExtractor} in the dual-image mode and adopted 
F356W-based match-aperture photometry. We chose the F356W band as the detection
band for the following reasons:  
(1) the F356W images are usually the deepest across all bands;
(2) the PSF in this band is sufficiently large, which
ensures that the apertures properly determined in this band include all source 
fluxes in any bluer bands;
(3) the F356W images present a notably cleaner background than those in the
SW bands.

    The source extraction was done by applying a $5\times5$ Gaussian 
convolution filter with a full width at half maximum (FWHM) of 2 pixels. From 
the weight images produced by the JWST data reduction pipeline, we derived the 
``root mean square'' (rms) maps using the \texttt{astroRMS} routine
\footnote{See \url{https://github.com/mmechtley/astroRMS}; we modified the
routine slightly by adopting a better source masking functionality.},
which calculates the auto-correlation of the science image pixels due to
the drizzling process that should be applied to scale the weight images. These
RMS maps were used for both the source detection and the estimate of 
photometric errors. 
The detection and analysis thresholds were set to 1.0 in \texttt{SExtractor}. 
We adopted the {\tt MAG\_ISO} magnitudes and only retained the sources with 
signal-to-noise ratio (S/N) of at least 5 and ISOAREA of at least 10 pixels in 
the F356W image. 

\subsection{Photometry for spectral energy distributions}

   When constructing SEDs, a common practice is to use photometry on 
PSF-matched images, i.e., the images in different 
bands are all convolved to have the same PSF size as in the band that has the 
largest PSF. Our SED analysis would involve both the NIRCam and the MIRI data, 
and PSF-matched photometry across all bands would not be appropriate in this 
case: smearing the NIRCam data (PSF FWHM 0\farcs030 to 0\farcs145 from F090W 
to F444W) to the coarsest MIRI resolution (F2550W PSF FWHM of 
$\sim$0\farcs803) would blend many unrelated neighbors in the NIRCam images 
and corrupt the NIRCam photometry.

   As the best compromise, we took a hybrid approach. In the NIRCam 
wavelengths, we prepared another set of NIRCam photometry done on the PSF-
matched NIRCam images. For each field, we convolved the images in the bluer 
bands to the PSF size of the F444W image. The PSFs were derived using isolated 
stars in the field, following the procedure in \citet[][]{LY2022}.
\texttt{SExtractor} was again run in the dual-image mode, and this time
the F444W image was used as the basis. We again adopted the {\tt MAG\_ISO} 
magnitudes to calculate the colors. A common practice would be to scale up the 
{\tt MAG\_ISO}-based SED to the ``total flux'' SED by adding the difference 
between the {\tt MAG\_ISO} and the {\tt MAG\_AUTO} magnitude in F444W band. 
{However, a considerable fraction of our objects have close neighbors that would contaminate the {\tt MAG\_AUTO} 
measurements, e.g., {\tt f115d\_brt\_cosmos\_344} and {\tt f150d\_brt\_cosmos\_093} shown in Figure~\ref{fig:dropouts}.}
Therefore, we chose not to apply such a correction. For isolated, bright 
($m_{444}<26.5$~mag) sources, we found that the differences between the 
\texttt{MAG\_ISO} and the \texttt{MAG\_AUTO} magnitudes were under 0.05~mag, 
and the impact of omitting such a correction would only be marginal to the SED 
analysis. 

   The MIRI PSF size varies greatly in different bands, which makes the 
PSF-matched photometry also inappropriate among the MIRI images. Therefore, we
ran \texttt{SExtractor} in the single-image mode for each MIRI band and 
adopted the {\tt MAG\_ISO} magnitudes. Similar to what was mentioned above for the 
NIRCam images, we also found that the difference between the {\tt MAG\_ISO} 
and the \texttt{MAG\_AUTO} magnitudes were under 0.05~mag for bright, isolated
sources in the MIRI images.

   Lastly, we note that we treated any photometry with S/N~$<2$ as 
non-detection and adopted the 2~$\sigma$ upper limit. Such limits were 
measured on the rms maps at the source locations using circular apertures of 
sizes equivalent to those of the {\tt MAG\_ISO} apertures.

     For simplicity, hereafter we denote the magnitudes in the HST/ACS  
F435W, F606W, F775W, F814W, and F850LP bands as $m_{435}$, $m_{606}$, 
$m_{775}$, $m_{814}$, $m_{850}$, respectively, those in the JWST/NIRCam F090W, 
F115W, F150W, F200W, F277W, F335M, F356W, F410M, and F444W bands as
$m_{090}$, $m_{115}$, $m_{150}$, $m_{200}$, $m_{277}$, $m_{335}$, $m_{356}$, 
$m_{410}$, and $m_{444}$, respectively, and those in the JWST/MIRI F560W, 
F770W, F1000W, F1280W, F1500W, F1800W, F2100W, and F2550W as $m_{560}$, 
$m_{770}$, $m_{1000}$, $m_{1280}$, $m_{1500}$, $m_{1800}$, $m_{2100}$, and 
$m_{2550}$, respectively. 

\section{Selection of Very Bright Dropouts}{\label{dropouts}}

    As mentioned in Section 3.1, the dropout selection was done using the 
NIRCam photometry based on the non-PSF-matched images. This approach has 
multiple advantages over using the PSF-matched images, such as no artificially 
introduced blending problem, less chance of misidentifying a broadened 
artifact as a source, more accurate S/N assessment, etc. The potential caveat 
of biased color indices using non-PSF-matched images has only a marginal 
impact here for two reasons. First, our objects would be dropouts from the 
bands bluer than F356W (the vast majority being dropouts from the SW bands), 
and the {\tt MAG\_ISO} aperture defined in F356W is large enough to include 
most (if not all) of the light in these bands in the first place. Second, the 
break amplitude is determined by the magnitudes in two adjacent bands whose 
PSF sizes are close, and the small difference in the fraction of light 
enclosed by the adopted aperture would only smear the redshift selection 
function negligibly.

    Similar to \citet[][]{Yan2023_smacs, Yan_highz_23}, we adopted the
dropout selection criteria as follows.

     (1) \emph{S/N $\geq5$ in the shift-in band.} The shift-in band is the 
redder band adjacent to the drop-out band. This criterion is to ensure the 
detection in the shift-in band and the robustness of the measured dropout 
amplitude. 

     (2) \emph{Dropout amplitude $\geq0.8$ mag.} The dropout amplitude is the 
color index between the drop-out and the shift-in bands. As mentioned above, 
when a source has $S/N<2$ in the drop-out band, its magnitude is substituted 
with the 2~$\sigma$ upper limit. This amplitude is chosen because the Lyman-
break signature shifted halfway out of the dropout band would create a color 
index of $\sim$0.75~mag between the dropout and shift-in bands for a flat 
spectrum in $f_\nu$.

    (3) \emph{S/N $\geq5$ in at least one more band redder than the
shift-in band.} All the retained sources have S/N $\geq 5$ in F356W and the 
shift-in band, and this additional requirement further ensures the reliability 
of the detections.

    (4) \emph{S/N $<2$ in all bands bluer than the drop-out band.} A legitimate
candidate should not be detected in these ``veto'' bands. This is
the most important criterion that distinguishes the dropout selection and the ERO
selection, whereas the latter does not have such a requirement. 

\begin{figure*}[hbt!]
    \centering
    \includegraphics[width=0.9\textwidth]{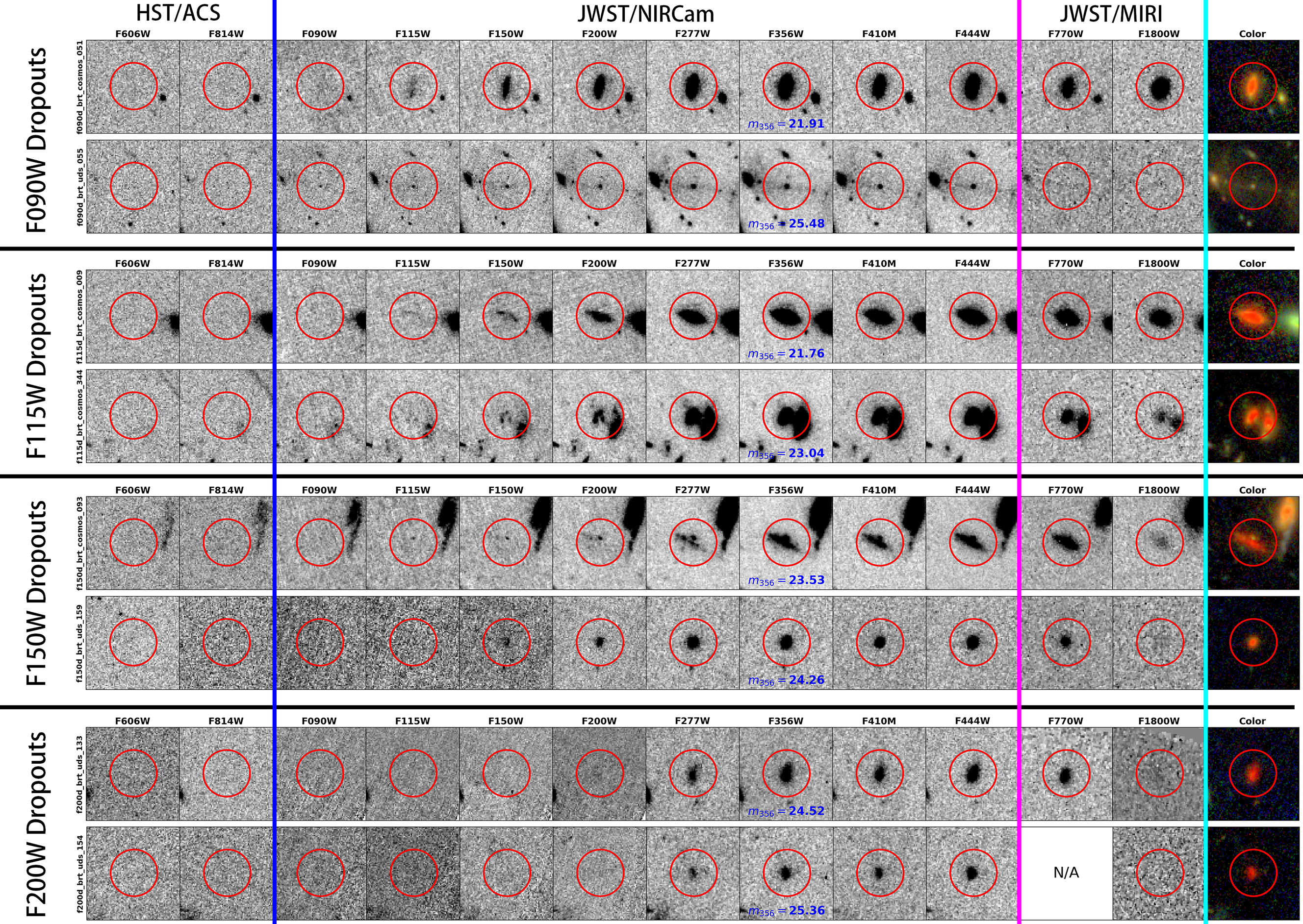}
    \caption{Image stamps of example very bright dropouts in F090W, F115W, F150W, 
    and F200W, arranged from top to bottom. Two example objects are shown for
    each group. The stamps are 2\arcsec$\times$2\arcsec\ in size and are oriented
    with north being up and east being left. The images are from the HST ACS, 
    JWST NIRCam and JWST MIRI, with the passbands as noted. Most of the very 
    bright dropouts are either disk-like ($\sim$40\%) or compact ($\sim$45\%) in 
    morphology in F356W, and one each is shown for the F090W, F150W and F200W 
    dropouts. The example F115W dropouts include a disk-like object and an 
    irregular object. Only $\sim$15\% of the very bright dropouts have irregular 
    morphology. 
    {The color stamps in the last column are constructed based on the NIRCam images, 
    using F090W + F115W + F150W as blue, F200W + F277W as green, and F356W + F410M + F444W as red. 
    }
    }
    \label{fig:dropouts}
\end{figure*}

     The candidates thus selected were then visually examined in all bands 
to ensure that they are real sources and are indeed invisible in the
veto bands. In this work, we only study very bright F090W and F115W dropouts with
$m_{356}\leq25.1$~mag and F150W, F200W, and F277W dropouts with 
$m_{356}\leq26.0$~mag. 
The objects fainter than these limits are considered ``normal'' and
are discussed in \citet[][]{Yan_highz_23} (for the F150W, F200W and F277W 
dropouts) and in Sun \& Yan (in preparation, for the F090W and F115W 
dropouts), respectively. In total, we have found 300 bright dropouts, 137 of 
which are covered by at least one MIRI band. We focus on these 
137 dropouts that have MIRI photometry (with detections or upper limits), 
which form our main sample and will be discussed next.
The other 163 objects form the supplement sample; for the sake of completeness,
these objects are listed in Table~\ref{tab:suppl}.
In the main sample, there are 59 dropouts in F090W, 36 in F115W, 37 in F150W, 5
in F200W, and none in F277W. Figure \ref{fig:dropouts} shows the image stamps of 
two objects in each group as examples. 
{The number of objects in the main sample in each field and each
passband are listed in Table~\ref{tab:dropouts}. }
If the dropout effect is due to the Lyman 
break, the nominal redshift ranges of the dropouts in F090W, F115W, F150W, and 
F200W are $z\approx 6.4$--8.4, 8.4--11.3, 11.3--15.4, 15.4--21.8, respectively.
The aforementioned brightness criteria roughly correspond to $M\lesssim -22$~mag
in rest frame U and B band, respectively.

{Moreover, we assessed the morphology of each source by visual inspection of the NIRCam images. }    
Interestingly, most of these objects are either very compact or have
disk-like morphology. Out of the 300 bright dropouts, $\sim$48\% are compact 
sources, $\sim$38\% are disk-like, $\sim$2\% are elliptical, 
and $\sim$12\% are irregular in shape.
In the pre-JWST era, one would put the disk-like objects to low redshifts  
because common wisdom has been that disk galaxies cannot be formed
so early in time. However, the JWST observations over the past two years have 
found a large number of candidate stellar disks at $z>2$ and up to $z\approx 8$
\citep[e.g.,][]{Fudamoto2022, Ferreira2022, Ferreira2023, Nelson2023a, 
Jacobs2023, Robertson2023b, Kuhn2024, YSL2024}, which suggests  
an early formation of stellar disks. Therefore, we chose not to make the judgment
based on morphology, and all objects were analyzed in the same way.

\begin{figure}
    \centering
    \includegraphics[width=0.95\linewidth]{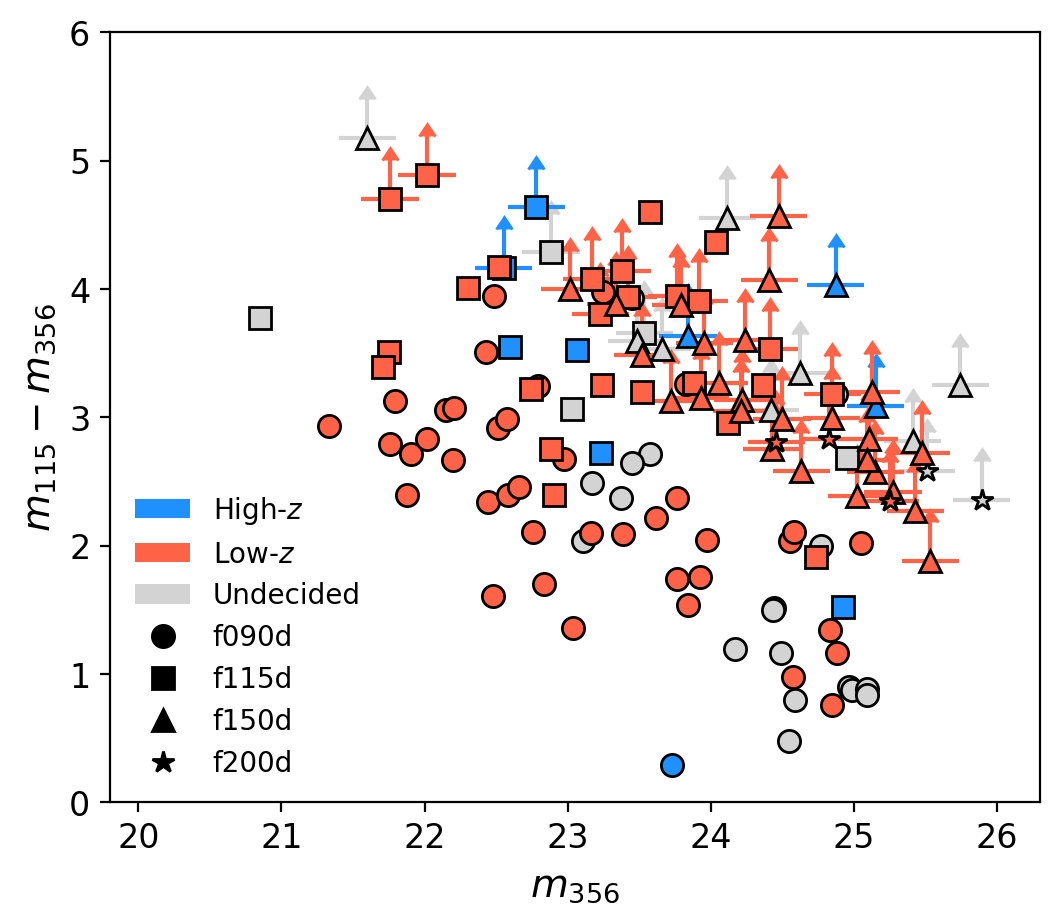}
    \caption{Observed $m_{115}-m_{356}$ color versus $m_{356}$ of the very bright
    dropouts in the main sample. The F090W, F115W, F150W, and F200W dropouts are 
    represented by circles, squares, triangles, and stars, respectively. The ones
    with upward arrows indicate the lower limits of the color if the objects
    have ${\rm S/N}<2.0$ in F115W, and the color lower limits are calculated 
    using the 2~$\sigma$ detection upper limits in this band. Adopting 
    $m_{115}-m_{356}>2.0$~mag as the fiducial criterion for EROs, the
    vast majority (81\%) of the very bright dropouts shown here would be
    selected. The objects in the ``High-$z$'', ``Low-$z$'', and ``Undecided'' 
    categories (see Section 5.2 for details) are shown in blue, red, and gray, 
    respectively.
    }
    \label{fig:115-356_color}
\end{figure}

    To demonstrate how the bright dropouts are qualified as EROs,
Figure~\ref{fig:115-356_color} shows the $m_{115}-m_{356}$ color distribution
as a function of $m_{356}$ for the 137 objects in our main sample, using different
symbols for the dropouts from different bands. Even when adopting a very red 
color of $m_{115}-m_{356}>2.0$~mag as the fiducial criterion for EROs, the vast 
majority (81\%) of our bright dropouts would be selected.

\begin{table*}[hbt!]
    \small
    \raggedright
    \begin{tabular}{ccccccc}
       Dropout Band & F090W & F115W & F150W & F200W & F277W & Total \\ \hline
        {\bf COSMOS} & 29 (69)& 10 (16)& 15 (24)& 1 (1) & 0 & 55 (110)\\
        High-z T1/T2 & 0/1& 0/1 & 2/0 & 0/0 & 0 & 2/2\\
        Low-z T1/T2 & 7/11& 2/5& 7/3& 0/1& 0 & 16/20\\ 
        Undecided & 10 & 2 & 3 & 0 & 0 & 15 \\
        \hline
       
       {\bf UDS} & 21 (27)& 20 (49)& 17 (35)& 4 (7)& 0 & 62 (118)\\  
       High-z T1/T2 & 0/0 & 0/4 & 0/0 & 0/0 & 0 & 0/4 \\ 
       Low-z T1/T2 & 10/7& 5/9& 9/5& 0/2& 0 & 24/23\\ 
       Undecided & 4 & 2 & 3 & 2 & 0 & 11 \\
       \hline
       
       {\bf CEERS} & / & 6 (37)& 5 (13)& 0 (2)& 0 & 11 (52)\\
       High-z T1/T2 & / & 0/1& 1/0 & 0 & 0 & 1/1\\ 
       Low-z T1/T2 & / & 2/2& 0/2& 0 & 0 & 2/4\\ 
       Undecided & / & 1 & 2 & 0 & 0 & 3 \\
       \hline
       
       {\bf GOODS-S} & 9 (12)& 0 (1)& 0 (6)& 0 (1)& 0 & 9 (20)\\
       High-z T1/T2 & 0/0 & 0/0 & 0 & 0 & 0 & 0/0 \\ 
       Low-z T1/T2 & 1/3& 0/0& 0 & 0 & 0 & 1/3\\ 
       Undecided & 5 & 0 & 0 & 0 & 0 & 5 \\
       \hline
       
       {\bf Total} & 59 (108)& 36 (103)& 37 (78)& 5 (11)& 0 & 137 (300)\\ 
    \end{tabular}
    \caption{Statistics of very bright dropouts in each field. For each field, 
    the first row gives the total numbers of bright dropouts in the main sample
    (i.e., objects covered by at least one MIRI band), with the numbers in 
    parentheses representing the total in the whole sample.  
    The second and the third rows show the numbers of ``High-$z$'' and 
    ``Low-$z$'' objects in T1 and T2 (separated by ``/''), respectively. The
    fourth row is the number of ``Undecided'' objects. See Section 5.2 for details.
    } 
    \label{tab:dropouts}
\end{table*}

\section{SED Analysis}

    To better understand the nature of these bright dropouts, we analyzed the SEDs 
of the objects in the main sample. The SEDs were constructed using the photometry 
in both NIRCam and MIRI as detailed in Section 3.2, which was tailored for this
purpose. Following the usual practice, we added 0.05~mag in quadrature to the 
reported photometric errors to account for the possible systematics, e.g., the 
offsets between the NIRCam and the MIRI photometry due to the different adopted 
methods.

\subsection{Methods and procedures}\label{sec:sedtools}

    We fitted the SEDs using three different tools, 
namely, Le Phare (version 2.2; \citealt{Arnouts1999,Ilbert2006}), 
EAZY (eazy-py version 0.6.4; \citealt{Brammer2008}), 
and CIGALE (version 2022.1; \citealt{Boquien19_cigale}). A major goal was to
obtain their photometric redshifts ($z_{\rm phot}$), which were allowed to vary
between $z=0$ to 25 in the fitting process.
The settings for each tool are as follows.  

$\bullet$ {\it Le Phare}: we constructed the templates using the population 
synthesis models of \citet[][hereafter ``BC03'']{Bruzual2003} and the initial 
mass function (IMF) of \citet{Chabrier2003}. We adopted an exponentially 
declining star formation history (SFH), i.e., SFR~$\propto e^{-t/\tau}$, where
$\tau$ ranges from 0 to 13 Gyr. We applied Calzetti's extinction law 
\citep[][]{Calzetti1994, Calzetti2001} with E(B-V) ranging from 0 to 1~mag at
a step size of 0.1~mag. The option to include the contribution from emission 
lines was turned on.

$\bullet$ {\it EAZY}: we used the template set {\tt GALSEDATLAS} of 
\cite{Brown14}, which was retrieved from MAST
\footnote{\url{https://archive.stsci.edu/hlsp/galsedatlas}}.
This set includes the spectra of 129 nearby galaxies of different types, which 
cover the UV-to-mid-IR wavelength range. Presumably, using this set of 
templates would favor the fitting at low redshifts.
{We note that we did not use the template sets optimized for fitting $z\gtrsim8$ galaxies \citep[e.g.,][]{Larson2023}.
Given the expectation that a significant fraction of the brightest dropouts would be EROs at low redshifts, 
using the {\tt GALSEDATLAS} templates helps exclude high-$z$ false positives as rigorously as possible. 
}

$\bullet$ {\it CIGALE}: we adopted a grid of CIGALE templates that include
those using a delayed $\tau$ model {in the form of SFR~$\propto te^{-t/\tau}$}
with $0.01<\tau\leq13$~Gyr, a recent 
starburst, and the simple stellar population (SSP) models (i.e., single
bursts) from BC03 assuming Chabrier IMF; we fixed the metallicity to 
$Z=0.02$. We set the nebular emission contribution with $-4<\log U<-1$ at
a step size of 0.5. We also adopted a modified Calzetti's extinction law 
under the {\tt dust\_modified\_starburst} module with the color excess of 
nebular gas E(B-V)$_g$ ranging from 0 to 3~mag at a step of 0.2~mag, and a 
fixed multiplication factor of 0.44 to apply on E(B-V)$_g$ to calculate the 
stellar continuum attenuation E(B-V)$_s$. 
There are a lot of cases where the bright dropouts have enhanced emission in
the central region increasing with wavelength, which could be caused by AGN.
To investigate this probability, we included the {\tt skirtor2016} AGN models 
from \cite{Stalevski12_agn,Stalevski16_agn}. We varied the AGN fraction 
(hereafter $f_{\rm AGN}$), i.e., the AGN contribution to $L_{IR}$, from 0 to 1 
at a step size of 0.2. We set the viewing angles at either $30^\circ$ or 
$70^\circ$ for Type 1 or 2 AGNs, respectively. 
We noticed that the program would run into errors when $f_{\rm AGN}=1$, 
and we resolved this by setting the maximum $f_{\rm AGN}$ to 0.999
\footnote{In the cases of CIGALE returning $f_{\rm AGN}=0.999$, they are reported as $f_{\rm AGN}=1.0$.} 
For the objects that do not show a compact central region at any passbands (in 
other words, their optical-to-IR emission is not likely due to AGN), we fixed 
their $f_{\rm AGN}=0$.

\begin{figure*}[hbt!]
    \centering
    \includegraphics[width=0.95\textwidth]{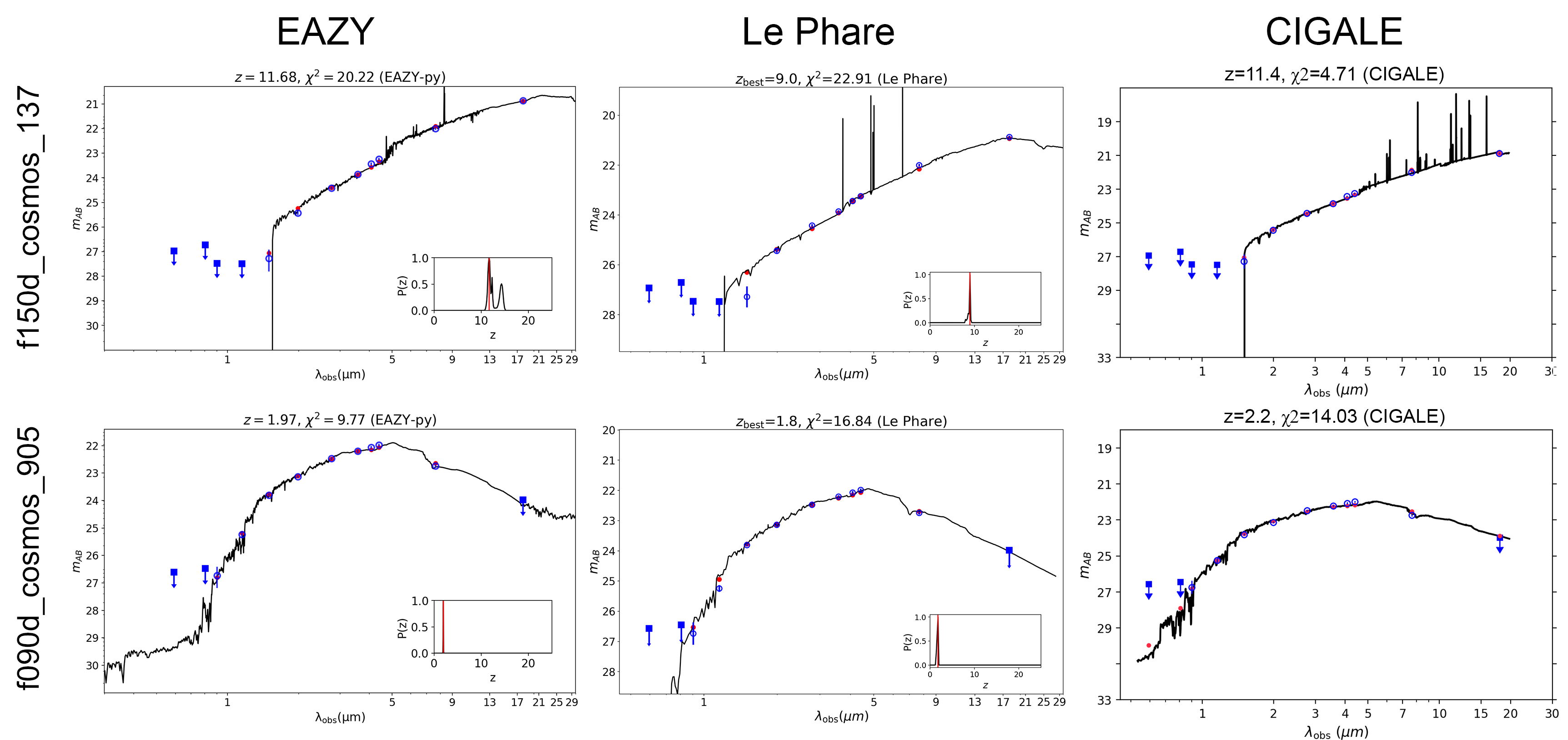}
    \caption{Examples of SED fitting results on one object in the ``High-$z$''
    category (top panel) and one in the ``Low-$z$'' category (bottom panel).
    The blue symbols represent the observed values, and the curves are the
    spectra of the best-fit models. The red symbols are the synthesized magnitudes
    derived from the best-fit models. The insets show the probability distribution
    function of $z_{\rm phot}$. The quoted $z_{\rm phot}$ value on top of each panel
    is from the shown best-fit model, which is slightly different from the
    adopted value (see Table~\ref{tab:t1highz} for further explanation.)}
    \label{fig:sed_example}
\end{figure*}

    Examples of SED fitting results from these three runs are provided in 
Figure \ref{fig:sed_example}. We note on the treatment of the upper limits.  
For Le Phare, we set the magnitude error to $-1$ when a given band has the
upper limit imposed, and the routine rejected any fits that violated the upper
limit. For EAZY, we used the modified code as described in
\citet[][]{Yan2023_smacs, Yan_highz_23} to achieve this functionality. 
For CIGALE, we set the fluxes to the upper limit 
and the corresponding flux errors to $-1$ times the upper limit. 

\subsection{Categorizing Bright Dropouts} \label{sec:sorting}

    We divided the bright dropouts in the main sample into three categories 
based on their $z_{\rm phot}$, namely, ``High-$z$'', ``Low-$z$'', and 
``Undecided''. The ``High-$z$'' category consists of objects for which at least 
two SED fitting tools (among the three) consistently derive 
$z_{\rm phot}\geq 6.0$. Similarly, the ``Low-$z$'' category consists of objects 
with consistent $z_{\rm phot}< 6.0$ from at least two fitting tools. In both 
categories, we further ranked the objects into ``Tier 1'' (T1) and ``Tier 2'' 
(T2) depending on the goodness of fits and the consistency of the results. For 
practical purpose, we deemed the fits with raw $\chi^2\leq100$ as ``good'' fits. 
For the ``High-$z$'' category, if an object has good fits and 
$z_{\rm phot}\geq 6.0$ from all three tools, it was put in T1; if it has good fits 
and $z_{\rm phot}\geq 6.0$ from only two tools, or if it has $z_{\rm phot}\geq 6.0$ 
from all three tools but the fits are not always good, it was placed in T2. The
ranking for the ``Low-$z$'' category was done similarly, with a slightly more
stringent requirement on the consistency of $z_{\rm phot}$: if an object has good 
fits and consistent $z_{\rm phot}$ from all three tools so that the differences 
$\Delta z_{\rm phot}<1.0$, it was put in T1; if it has good fits and consistent 
$z_{\rm phot}$ from only two tools, or if it has consistent $z_{\rm phot}$ from all
three tools but the fits are not always good, it was placed in T2. Finally, all 
objects that were not in either the ``High-$z$'' or the ``Low-$z$'' category as
ranked were assigned to the ``Undecided'' category.

$\bullet$ {\bf High-$z$: }
We identified 10 high-$z$ candidates from the sample, among which 3 are in T1 
and 7 are in T2. They are listed in Table \ref{tab:t1highz} and 
\ref{tab:t2highz}, respectively, along with their physical properties derived 
from the SED analysis. 

\begin{table*}[hbt!]
    \caption{Tier 1 ``High-$z$'' Objects} \label{tab:t1highz}
    \resizebox{\textwidth}{!}{
    \begin{tabular}{lcccccccccccccc} \hline
       SID & R.A. & Decl. & $m_{\rm in}$ & $m_{356}$ & $z_{lp}$ & $z_{ez}$ & $z_{cg}$ & $\log(M_*)$ [$M_\odot$] & $f_{AGN}$ & $E(B-V)$ & Age (Myr) & $\log$(SFR) [$M_\odot$/yr] & Morph \\ \hline
       f150d\_brt\_ceers\_051 &  215.0313294 & 52.9171141 & 26.57 & 24.89 & $12.22_{-0.28}^{+0.24}$ & $14.57_{-0.01}^{+0.01}$ & $14.38_{-0.73}^{+0.62}$ & $11.47_{-0.09}^{+0.06}$/$11.90_{-11.90}^{+0.37}$ & .../$0.62_{-0.27}^{+0.27}$ & 0.7/$0.82_{-0.26}^{+0.41}$ & $34_{-19}^{+78}$/$55_{-55}^{+74}$ & $5.36_{-0.51}^{+0.40}$/$4.77_{-4.77}^{+0.34}$ & c \\ 
       f150d\_brt\_cosmos\_093 & 150.0651062 & 2.2636221 & 27.60 & 23.53 & $10.66_{-0.23}^{+0.25}$ & $11.06_{-0.02}^{+0.02}$ & $9.95_{-0.18}^{+0.13}$ & $10.86_{-0.05}^{+0.20}$/$10.99_{-0.08}^{+0.07}$ & .../0 & 0.5/$0.59_{-0.05}^{+0.05}$ & $10_{-1}^{+1}$/$11_{-11}^{+11}$ & $3.91_{-0.11}^{+0.29}$/$4.29_{-0.09}^{+0.08}$ & d \\ 
       f150d\_brt\_cosmos\_137 & 150.1089402 & 2.2936631 & 25.49 & 23.86 & $8.81_{-0.53}^{+0.27}$ & $11.68_{-0.22}^{+0.22}$ & $10.75_{-2.31}^{+2.31}$ & $11.11_{-0.05}^{+0.04}$/$11.17_{-1.17}^{+0.29}$ & .../$0.35_{-0.24}^{+0.24}$ & 0.7/$0.60_{-0.21}^{+0.21}$ & $11_{-1}^{+3}$/$14_{-14}^{+18}$ & $4.07_{-0.07}^{+0.09}$/$4.14_{-4.14}^{+0.40}$ & c \\ \hline 
    \end{tabular}
    }
    \tablecomments{The brightness of an object is indicated using 
    {the shift-in band magnitude ($m_{\rm in}$) and} $m_{356}$. 
    The photometric redshifts from Le Phare, EAZY, and CIGALE are 
    listed as $z_{lp}$, $z_{ez}$, and $z_{cg}$, respectively.
    The stellar mass ($M_*$), reddening ($E(B-V)$) and star formation rate (SFR) 
    estimates are from the Le Phare and the CIGALE runs but are not 
    available in the EAZY runs, and therefore only two sets of values are
    quoted (separated by ``/''). The AGN fraction ($f_{\rm AGN}$) estimates are 
    only available from CIGALE. 
    {For Le Phare and EAZY, the photometric redshifts are the mean values 
    weighted by the probability distribution function $P(z)$.
    For CIGALE, $z_{cg}$, along with  stellar mass, $f_{\rm AGN}$, $E(B-V)$, Age, and SFR are all the 50th percentile values, too, and the errors indicate the 16th and 84th percentile values. 
    }
    The morphological classifications are given under
    the last column, where ``c'' stands for ``compact'' and ``d'' stands for
    ``disk-like''.
    } 
\end{table*} 

\begin{table*}[hbt!]
    \caption{Tier 2 ``High-$z$'' Objects} \label{tab:t2highz}
    \resizebox{\textwidth}{!}{
    \begin{tabular}{lcccccccccccccc} \hline
       SID & R.A. & Decl. & $m_{\rm in}$ & $m_{356}$ & $z_{lp}$ & $z_{ez}$ & $z_{cg}$ & $\log(M_*)$ [$M_\odot$] & $f_{AGN}$ & $E(B-V)$ & Age (Myr) & $\log$(SFR) [$M_\odot$/yr] & Morph \\ \hline 
        f115d\_brt\_ceers\_062 & 215.0354323 & 52.8906847 & 24.95 & 24.92 & $8.80_{-0.14}^{+0.14}$ & $8.88_{-0.01}^{+0.01}$ & $8.95_{-0.12}^{+0.12}$ & $9.26_{-0.04}^{+0.04}$/$8.94_{-0.11}^{+0.09}$ & .../$0.32_{-0.29}^{+0.29}$ & 0.2/$0.09_{-0.02}^{+0.02}$ & $11_{-1}^{+3}$/$15_{-15}^{+44}$ & $2.60_{-0.07}^{+0.08}$/$2.19_{-0.09}^{+0.09}$ & c \\ 
        f090d\_brt\_cosmos\_663 & 150.0634749 & 2.3552383 & 24.03 & 23.56 & $6.16_{-0.16}^{+0.16}$ & $6.88_{-0.00}^{+0.00}$ & $6.78_{-0.09}^{+0.09}$ & -/$9.04_{-0.11}^{+0.09}$ & -/$0.43_{-0.36}^{+0.36}$ & -/$0.01_{-0.01}^{+0.03}$ & -/$21_{-12}^{+12}$ & -/$2.11_{-0.11}^{+0.11}$ & c \\ 
        f115d\_brt\_cosmos\_270 & 150.0985778 & 2.3208768 & 25.45 & 22.75 & $7.93_{-0.83}^{+0.20}$ & $6.70_{-0.01}^{+0.01}$ & $2.86_{-0.78}^{+0.78}$ & $11.50_{-0.09}^{+0.05}$/- & .../- & 0.7/- & $111_{-8}^{+10}$/- & $4.81_{-0.18}^{+0.09}$/- & d \\ 
        f115d\_brt\_uds\_089 & 34.2669263 & -5.2947891 & 25.05 & 23.25 & $8.37_{-0.28}^{+0.23}$ & $2.64_{-0.07}^{+0.07}$ & $7.62_{-1.15}^{+1.15}$ & $10.76_{-0.05}^{+0.04}$/$10.69_{-0.75}^{+0.26}$ & .../$0.44_{-0.29}^{+0.29}$ & 0.5/$0.52_{-0.23}^{+0.23}$ & $12_{-2}^{+1}$/$257_{-257}^{+327}$ & $3.69_{-0.08}^{+0.08}$/$3.09_{-3.09}^{+0.52}$ & c \\
        f115d\_brt\_uds\_245 & 34.4716256 & -5.256956 & 24.38 & 22.62 & $8.78_{-0.14}^{+0.15}$ & $2.46_{-0.02}^{+0.02}$ & $8.55_{-1.47}^{+1.47}$ & $10.99_{-0.03}^{+0.04}$/$10.49_{-10.49}^{+0.41}$ & .../$0.72_{-0.23}^{+0.23}$ & 0.5/$0.48_{-0.34}^{+0.34}$ & $10_{-1}^{+1}$/$166_{-166}^{+225}$ & $4.04_{-0.09}^{+0.08}$/$2.50_{-2.50}^{+0.49}$ & irr \\
        f115d\_brt\_uds\_647 & 34.2421684 & -5.1472847 & 25.06 & 23.06 & $8.66_{-0.20}^{+0.17}$ & $2.54_{-0.02}^{+0.02}$ & $7.54_{-2.24}^{+2.24}$ & $10.80_{-0.04}^{+0.04}$/$10.74_{-0.24}^{+0.15}$ & .../0 & 0.5/$0.54_{-0.06}^{+0.06}$ & $10_{-1}^{+1}$/$160_{-160}^{+476}$ & $3.79_{-0.07}^{+0.07}$/$3.98_{-0.27}^{+0.17}$ & c \\
        f115d\_brt\_uds\_754 & 34.3546998 & -5.1097457 & 24.96 & 22.57 & $5.60_{-0.14}^{+0.14}$ & $9.27_{-0.02}^{+0.06}$ & $9.91_{-1.84}^{+1.84}$ & $11.17_{-0.05}^{+0.08}$/$11.52_{-0.27}^{+0.17}$ & .../0 & 0.9/$0.60_{-0.09}^{+0.09}$ & $10_{-1}^{+1}$/$27_{-27}^{+84}$ & $4.12_{-0.09}^{+0.07}$/$4.65_{-0.34}^{+0.19}$ & irr \\ \hline 
    \end{tabular}
    }
    \tablecomments{
    Similar to Table \ref{tab:t1highz} but for the Tier 2 objects in the
    ``High-$z$'' category. Some of the parameters derived by Le Phare are not
    available because of the very bad fits. 
    {The redshift solution by CIGALE for {\tt f115d\_brt\_cosmos\_270} is inconsistent with the other two tools, 
    and its derived parameters are therefore not included for comparison. }
    The morphological type ``irr'' stands for ``irregular''.
    } 
\end{table*}

$\bullet$ {\bf Low-$z$: }
There are 43 T1 and 50 T2 objects in this category, which are listed in
Tables~\ref{tab:t1lowz} and \ref{tab:t2lowz}, respectively. 
Among the 93 objects, the vast majority of them have $z_{\rm phot}=1$--4. 


\begin{table*}[hbt!]
    \caption{Tier 1 ``Low-$z$'' Objects} \label{tab:t1lowz}
    \resizebox{\textwidth}{!}{
    \begin{tabular}{lccccccccccccc} \hline
       SID & R.A. & Decl. & $m_{\rm in}$ & $m_{356}$ & $z_{lp}$ & $z_{ez}$ & $z_{cg}$ & $\log(M_*)$ [$M_\odot$] & $f_{AGN}$ & $E(B-V)$ & Age (Gyr) & $\log$(SFR) [$M_\odot$/yr] & Morph \\ \hline
        f115d\_brt\_ceers\_146 & 215.041352 & 52.914094 & 25.16 & 22.53 & $1.80_{-0.15}^{+0.15}$ & $2.64_{-0.06}^{+0.06}$ & $1.69_{-0.41}^{+0.32}$ & $10.49_{-0.11}^{+0.08}$/$10.54_{-0.06}^{+0.01}$ & .../0 & 0.6/$0.68_{-0.18}^{+0.21}$ & $1.60_{-0.75}^{+0.59}$/$1.56_{-0.85}^{+1.47}$ & $0.88_{-0.19}^{+0.35}$/$-11.53_{-34.29}^{+10.98}$ & d \\
        
        f115d\_brt\_ceers\_309 & 214.9785887 & 52.9215485 & 24.88 & 23.29 & $2.62_{-0.15}^{+0.14}$ & $2.40_{-0.07}^{+0.07}$ & $2.50_{-0.47}^{+0.30}$ & $10.29_{-0.04}^{+0.04}$/$10.29_{-0.03}^{+0.01}$ & .../$0.21_{-0.21}^{+0.24}$ & 0.0/$0.06_{-0.05}^{+0.11}$ & $1.77_{-0.18}^{+0.18}$/$1.76_{-0.71}^{+0.82}$ & $-4.77_{-0.83}^{+2.77}$/$-38.01_{-23.87}^{+24.91}$ & c \\ 
        f090d\_brt\_cosmos\_296 & 150.0880207 & 2.2795606 & 25.84 & 23.64 & $1.46_{-0.17}^{+0.15}$ & $1.61_{-0.03}^{+0.03}$ & $1.62_{-0.22}^{+0.21}$ & $9.83_{-0.06}^{+0.05}$/$9.86_{-0.01}^{+0.01}$ & .../0 & 0.1/$0.04_{-0.03}^{+0.05}$ & $3.78_{-0.61}^{+0.67}$/$3.26_{-0.33}^{+0.42}$ & $-3.16_{-1.64}^{+1.26}$/$-26.62_{-74.26}^{+24.99}$ & e \\ 
        
        f090d\_brt\_cosmos\_364 & 150.0869802 & 2.289832 & 24.85 & 22.02 & $1.78_{-0.14}^{+0.15}$ & $1.60_{-0.01}^{+0.01}$ & $1.24_{-0.16}^{+0.19}$ & -/$10.16_{-0.01}^{+0.01}$ & .../0 & 0.6/$0.61_{-0.04}^{+0.04}$ & -/$3.34_{-1.05}^{+1.04}$ & -/$0.89_{-0.01}^{+0.01}$ & irr \\ 
        
        f090d\_brt\_cosmos\_543 & 150.1822751 & 2.3270705 & 26.04 & 22.80 & $2.54_{-0.14}^{+0.17}$ & $2.41_{-0.02}^{+0.02}$ & $2.48_{-0.20}^{+0.24}$ & $10.55_{-0.07}^{+0.07}$/$10.54_{-0.02}^{+0.01}$ & .../$0.44_{-0.35}^{+0.35}$ & 0.0/$0.04_{-0.03}^{+0.05}$ & $1.78_{-0.32}^{+0.26}$/$2.06_{-0.60}^{+0.55}$ & $-4.31_{-1.18}^{+3.15}$/$-43.07_{-28.63}^{+28.54}$ & c \\ 
        
        f090d\_brt\_cosmos\_589 & 150.0687834 & 2.342366 & 26.05 & 24.97 & $1.42_{-0.19}^{+0.17}$ & $1.59_{-0.06}^{+0.06}$ & $1.42_{-0.19}^{+0.21}$ & $9.03_{-0.53}^{+0.18}$/$9.23_{-0.01}^{+0.01}$ & .../0 & 0.2/$0.05_{-0.04}^{+0.07}$ & $2.61_{-2.48}^{+1.32}$/$3.35_{-0.49}^{+0.56}$ & $-0.38_{-0.36}^{+0.49}$/$-2.26_{-90.99}^{+0.78}$ & c \\ 
        
        f090d\_brt\_cosmos\_653 & 150.0649198 & 2.3573118 & 25.55 & 24.64 & $0.62_{-0.16}^{+0.17}$ & $1.04_{-0.03}^{+0.03}$ & $0.99_{-0.16}^{+0.19}$ & $7.88_{-0.06}^{+0.05}$/$8.92_{-0.01}^{+0.01}$ & .../0 & 1.0/$0.03_{-0.03}^{+0.04}$ & $0.01_{-0.00}^{+6.91}$/$2.04_{-0.32}^{+0.36}$ & $1.00_{-0.09}^{+0.08}$/$-17.26_{-45.45}^{+15.02}$ & d \\ 
        
        f090d\_brt\_cosmos\_676 & 150.0844722 & 2.3576783 & 24.40 & 23.10 & $1.20_{-0.14}^{+0.14}$ & $1.13_{-0.04}^{+0.04}$ & $0.99_{-0.16}^{+0.18}$ & $9.78_{-1.53}^{+0.10}$/$9.65_{-0.01}^{+0.01}$ & .../0 & 0.0/$0.03_{-0.03}^{+0.04}$ & $4.65_{-1.12}^{+0.77}$/$3.80_{-0.80}^{+1.03}$ & $2.83_{-6.74}^{+0.23}$/$-29.98_{-88.00}^{+28.13}$ & c \\ 
        
        f090d\_brt\_cosmos\_905 & 150.1387124 & 2.4440275 & 25.28 & 22.21 & $1.76_{-0.15}^{+0.16}$ & $1.96_{-0.01}^{+0.01}$ & $1.99_{-0.15}^{+0.19}$ & $10.60_{-0.06}^{+0.05}$/$10.59_{-0.01}^{+0.01}$ & .../$0.46_{-0.35}^{+0.35}$ & 0.1/$0.03_{-0.03}^{+0.04}$ & $3.42_{-0.31}^{+0.40}$/$3.05_{-0.46}^{+0.33}$ & $-5.05_{-0.10}^{+2.33}$/$-64.78_{-43.85}^{+42.81}$ & c \\ 
        
        f115d\_brt\_cosmos\_086 & 150.0838418 & 2.236936 & 25.81 & 24.84 & $1.41_{-0.15}^{+0.14}$ & $1.34_{-0.02}^{+0.02}$ & $1.21_{-0.17}^{+0.21}$ & $9.15_{-1.26}^{+0.13}$/$9.17_{-0.01}^{+0.01}$ & .../0 & 0.0/$0.04_{-0.04}^{+0.05}$ & $4.80_{-0.44}^{+0.51}$/$3.96_{-0.75}^{+0.57}$ & $2.83_{-0.09}^{+0.07}$/$-30.00_{-92.46}^{+27.73}$ & c \\ 
        
        f115d\_brt\_cosmos\_226 & 150.1113201 & 2.2988013 & 25.90 & 24.13 & $3.31_{-0.28}^{+0.21}$ & $2.80_{-0.06}^{+0.06}$ & $2.31_{-0.22}^{+0.23}$ & $9.70_{-0.19}^{+0.12}$/$9.99_{-0.01}^{+0.01}$ & .../0 & 0.8/$0.36_{-0.07}^{+0.06}$ & $1.05_{-1.01}^{+0.92}$/$0.81_{-0.39}^{+0.38}$ & $2.84_{-0.18}^{+0.17}$/$-6.12_{-16.43}^{+5.36}$ & d \\ 
        
        f150d\_brt\_cosmos\_010 & 150.0949689 & 2.1771805 & 25.12 & 24.09 & $3.11_{-0.26}^{+0.27}$ & $3.03_{-0.07}^{+0.07}$ & $3.41_{-0.48}^{+0.50}$ & $9.90_{-0.15}^{+0.14}$/$10.16_{-0.03}^{+0.03}$ & .../$0.40_{-0.35}^{+0.35}$ & 1.0/$0.05_{-0.04}^{+0.07}$ & $2.84_{-0.18}^{+0.17}$/$1.36_{-0.56}^{+0.60}$ & $3.22_{-0.17}^{+0.15}$/$-27.93_{-18.05}^{+18.43}$ & c \\ 
        
        f150d\_brt\_cosmos\_044 & 150.0797437 & 2.2273521 & 26.24 & 25.04 & $3.04_{-0.22}^{+0.25}$ & $3.55_{-0.11}^{+0.11}$ & $3.44_{-0.30}^{+0.29}$ & $9.47_{-0.23}^{+0.31}$/$10.09_{-0.03}^{+0.01}$ & .../0 & 0.6/$0.44_{-0.09}^{+0.08}$ & $0.04_{-0.03}^{+0.48}$/$0.54_{-0.46}^{+0.64}$ & $2.21_{-0.86}^{+0.11}$/$-0.42_{-9.31}^{+0.99}$ & c \\ 
        
        f150d\_brt\_cosmos\_160 & 150.0817351 & 2.3043475 & 26.69 & 25.13 & $1.71_{-0.19}^{+0.19}$ & $2.43_{-0.05}^{+0.05}$ & $1.52_{-0.43}^{+0.42}$ & $8.72_{-0.28}^{+0.59}$/$9.46_{-0.06}^{+0.02}$ & .../$0.14_{-0.14}^{+0.20}$ & 0.8/$0.77_{-0.21}^{+0.21}$ & $0.07_{-0.04}^{+1.24}$/$2.43_{-1.57}^{+1.04}$ & $0.83_{-0.59}^{+0.28}$/$0.03_{-0.60}^{+0.09}$ & c \\ 
        
        f150d\_brt\_cosmos\_384 & 150.176541 & 2.4599094 & 26.24 & 25.37 & $2.11_{-0.22}^{+0.22}$ & $3.17_{-0.25}^{+0.25}$ & $2.44_{-0.42}^{+0.66}$ & $8.64_{-0.12}^{+0.44}$/$9.53_{-0.02}^{+0.01}$ & .../0 & 0.7/$0.36_{-0.19}^{+0.17}$ & $0.03_{-0.02}^{+0.27}$/$1.18_{-0.65}^{+0.92}$ & $1.16_{-0.45}^{+0.39}$/$-8.06_{-26.51}^{+6.83}$ & c \\ 
        
        f150d\_brt\_cosmos\_389 & 150.1539149 & 2.4671568 & 26.68 & 25.33 & $3.40_{-0.41}^{+0.31}$ & $3.64_{-0.19}^{+0.19}$ & $3.74_{-0.47}^{+0.31}$ & $9.39_{-0.25}^{+0.24}$/$9.97_{-0.02}^{+0.01}$ & .../0 & 0.9/$0.25_{-0.08}^{+0.09}$ & $1.20_{-1.17}^{+0.67}$/$0.91_{-0.50}^{+0.51}$ & $2.47_{-0.60}^{+0.30}$/$-7.27_{-18.37}^{+6.42}$ & c \\ 
        
        f150d\_brt\_cosmos\_390 & 150.1586059 & 2.4683481 & 25.14 & 24.00 & $3.05_{-0.21}^{+0.21}$ & $3.58_{-0.02}^{+0.02}$ & $3.39_{-0.27}^{+0.28}$ & $9.67_{-0.13}^{+0.40}$/$10.49_{-0.02}^{+0.01}$ & .../0 & 0.9/$0.34_{-0.08}^{+0.06}$ & $0.01_{-0.00}^{+0.26}$/$0.81_{-0.36}^{+0.36}$ & $2.50_{-0.76}^{+0.23}$/$-2.53_{-18.95}^{+2.28}$ & d \\ 
        
        f150d\_brt\_cosmos\_400 & 150.149107 & 2.4827173 & 24.99 & 24.23 & $2.30_{-0.27}^{+0.28}$ & $2.29_{-0.10}^{+0.10}$ & $2.73_{-0.56}^{+0.46}$ & $9.87_{-0.68}^{+0.12}$/$10.04_{-0.03}^{+0.01}$ & .../$0.36_{-0.35}^{+0.35}$ & 1.0/$0.05_{-0.05}^{+0.11}$ & $2.28_{-0.68}^{+0.64}$/$1.85_{-0.91}^{+0.72}$ & $1.63_{-3.14}^{+1.69}$/$-38.80_{-25.68}^{+25.42}$ & c \\ 
        
        f090d\_brt\_jsmed\_191 & 53.0471741 & -27.8700154 & 27.22 & 22.83 & $3.19_{-0.18}^{+0.18}$ & $3.64_{-0.03}^{+0.03}$ & $2.88_{-0.32}^{+0.32}$ & $10.28_{-0.07}^{+0.14}$/$10.81_{-0.01}^{+0.01}$ & .../0 & 1.0/$0.67_{-0.10}^{+0.09}$ & $0.01_{-0.00}^{+0.01}$/$0.32_{-0.30}^{+0.56}$ & $3.26_{-0.21}^{+0.20}$/$0.01_{-5.69}^{+0.42}$ & irr \\ 
        
        f090d\_brt\_uds\_008 & 34.4023528 & -5.2907091 & 25.26 & 23.20 & $1.74_{-0.18}^{+0.17}$ & $1.74_{-0.02}^{+0.02}$ & $1.76_{-0.20}^{+0.22}$ & $10.06_{-0.09}^{+0.06}$/$10.12_{-0.01}^{+0.01}$ & .../$0.41_{-0.35}^{+0.35}$ & 0.1/$0.04_{-0.04}^{+0.06}$ & $2.42_{-0.67}^{+0.81}$/$2.85_{-0.82}^{+0.50}$ & $-0.32_{-0.33}^{+0.26}$/$-60.31_{-40.49}^{+40.19}$ & c \\ 
        
        f090d\_brt\_uds\_020 & 34.4040888 & -5.283555 & 24.87 & 22.23 & $1.80_{-0.14}^{+0.14}$ & $1.89_{-0.01}^{+0.01}$ & $1.93_{-0.16}^{+0.21}$ & $10.51_{-2.05}^{+0.07}$/$10.59_{-0.01}^{+0.01}$ & .../$0.37_{-0.36}^{+0.36}$ & 0.1/$0.05_{-0.04}^{+0.06}$ & $3.40_{-0.39}^{+0.44}$/$3.13_{-0.37}^{+0.28}$ & $3.32_{-5.99}^{+0.11}$/$-67.46_{-43.51}^{+45.00}$ & c \\ 
        
        f090d\_brt\_uds\_035 & 34.2395469 & -5.2770455 & 26.13 & 23.77 & $1.83_{-0.16}^{+0.14}$ & $1.82_{-0.04}^{+0.04}$ & $1.83_{-0.22}^{+0.19}$ & $9.85_{-0.06}^{+0.06}$/$9.94_{-0.01}^{+0.01}$ & .../$0.26_{-0.25}^{+0.25}$ & 0.1/$0.04_{-0.03}^{+0.05}$ & $1.89_{-0.29}^{+0.78}$/$2.73_{-0.82}^{+0.57}$ & $-1.88_{-3.33}^{+0.93}$/$-58.12_{-38.26}^{+37.98}$ & c \\ 
        
        f090d\_brt\_uds\_038 & 34.3312534 & -5.2762723 & 26.02 & 24.00 & $1.80_{-0.16}^{+0.15}$ & $2.10_{-0.02}^{+0.02}$ & $1.78_{-0.26}^{+0.23}$ & $9.62_{-0.11}^{+0.08}$/$9.76_{-0.04}^{+0.02}$ & .../0 & 0.4/$0.19_{-0.09}^{+0.08}$ & $1.21_{-0.88}^{+0.66}$/$1.53_{-0.72}^{+1.35}$ & $0.46_{-0.33}^{+0.16}$/$-12.07_{-33.76}^{+10.68}$ & irr \\ 
        
        f090d\_brt\_uds\_087 & 34.3935038 & -5.2608125 & 25.48 & 23.45 & $1.69_{-0.17}^{+0.17}$ & $1.75_{-0.05}^{+0.05}$ & $1.77_{-0.20}^{+0.21}$ & $9.93_{-0.07}^{+0.06}$/$10.03_{-0.01}^{+0.01}$ & .../$0.44_{-0.35}^{+0.35}$ & 0.1/$0.04_{-0.04}^{+0.05}$ & $2.40_{-0.59}^{+0.95}$/$2.91_{-0.88}^{+0.46}$ & $-2.22_{-2.20}^{+5.08}$/$-64.99_{-40.37}^{+42.87}$ & c \\ 
        
        f090d\_brt\_uds\_261 & 34.2702143 & -5.1423084 & 26.59 & 24.68 & $1.49_{-0.18}^{+0.17}$ & $1.49_{-0.01}^{+0.01}$ & $1.39_{-0.23}^{+0.22}$ & $9.33_{-1.29}^{+0.09}$/$9.27_{-0.01}^{+0.01}$ & .../$0.45_{-0.35}^{+0.35}$ & 0.0/$0.04_{-0.03}^{+0.05}$ & $4.02_{-0.54}^{+0.52}$/$3.56_{-0.52}^{+0.63}$ & $2.96_{-0.15}^{+0.16}$/$-76.32_{-50.79}^{+50.15}$ & c \\ 
        
        f090d\_brt\_uds\_289 & 34.3824186 & -5.1345838 & 25.94 & 22.44 & $2.39_{-0.15}^{+0.14}$ & $2.46_{-0.01}^{+0.01}$ & $2.43_{-0.23}^{+0.24}$ & $10.71_{-0.10}^{+0.08}$/$10.74_{-0.01}^{+0.01}$ & .../0 & 0.1/$0.14_{-0.06}^{+0.07}$ & $2.14_{-0.24}^{+0.28}$/$2.12_{-0.61}^{+0.55}$ & $-1.37_{-1.00}^{+0.35}$/$-16.48_{-46.90}^{+15.67}$ & irr \\ 
        
        f090d\_brt\_uds\_296 & 34.4194596 & -5.1337748 & 25.12 & 22.71 & $1.84_{-0.16}^{+0.14}$ & $2.10_{-0.04}^{+0.04}$ & $1.81_{-0.24}^{+0.21}$ & $10.21_{-0.05}^{+0.06}$/$10.35_{-0.02}^{+0.01}$ & .../0 & 0.4/$0.20_{-0.04}^{+0.04}$ & $0.59_{-0.15}^{+0.71}$/$2.14_{-0.75}^{+0.99}$ & $0.30_{-0.16}^{+0.25}$/$-16.74_{-46.87}^{+15.48}$ & c \\ 
        
        f090d\_brt\_uds\_310 & 34.5003733 & -5.1291902 & 25.50 & 23.74 & $1.16_{-0.59}^{+0.17}$ & $1.25_{-0.01}^{+0.01}$ & $1.08_{-0.21}^{+0.17}$ & $8.91_{-0.43}^{+0.41}$/$9.47_{-0.01}^{+0.01}$ & .../$0.49_{-0.34}^{+0.34}$ & 0.0/$0.03_{-0.03}^{+0.04}$ & $6.25_{-1.19}^{+1.68}$/$4.09_{-0.98}^{+0.80}$ & $2.17_{-6.50}^{+0.53}$/$-87.53_{-59.46}^{+58.22}$ & c \\ 
        
        f090d\_brt\_uds\_339 & 34.3837744 & -5.1194462 & 25.68 & 23.96 & $1.68_{-0.17}^{+0.16}$ & $1.64_{-0.05}^{+0.05}$ & $1.26_{-0.21}^{+0.23}$ & $9.70_{-0.08}^{+0.09}$/$9.48_{-0.02}^{+0.01}$ & .../$0.36_{-0.31}^{+0.31}$ & 0.1/$0.22_{-0.10}^{+0.08}$ & $2.21_{-0.41}^{+1.16}$/$1.84_{-1.01}^{+1.65}$ & $-1.87_{-1.56}^{+5.07}$/$-39.05_{-24.63}^{+24.95}$ & c \\ 
        
        f115d\_brt\_uds\_151 & 34.39082 & -5.2791092 & 25.15 & 23.55 & $1.71_{-0.33}^{+0.25}$ & $2.46_{-0.01}^{+0.01}$ & $2.44_{-0.16}^{+0.22}$ & $9.29_{-0.13}^{+0.45}$/$10.26_{-0.03}^{+0.01}$ & .../0 & 0.6/$0.03_{-0.03}^{+0.04}$ & $0.08_{-0.07}^{+0.20}$/$2.47_{-0.45}^{+0.34}$ & $2.13_{-1.03}^{+0.14}$/$-17.91_{-56.50}^{+16.64}$ & d \\ 
        
        f115d\_brt\_uds\_265 & 34.3571218 & -5.2531005 & 25.94 & 23.91 & $2.88_{-0.22}^{+0.18}$ & $3.03_{-0.03}^{+0.03}$ & $2.73_{-0.26}^{+0.26}$ & $10.05_{-0.31}^{+0.10}$/$10.30_{-0.05}^{+0.01}$ & .../0 & 0.5/$0.19_{-0.08}^{+0.08}$ & $0.10_{-0.03}^{+0.58}$/$1.59_{-0.65}^{+0.61}$ & $1.22_{-0.38}^{+1.45}$/$-12.77_{-33.83}^{+11.82}$ & d \\ 
        
        f115d\_brt\_uds\_685 & 34.3435201 & -5.1347536 & 26.08 & 23.40 & $3.17_{-0.16}^{+0.17}$ & $3.41_{-0.06}^{+0.06}$ & $2.66_{-0.32}^{+0.32}$ & $10.61_{-0.11}^{+0.14}$/$10.62_{-0.06}^{+0.01}$ & .../0 & 1.0/$0.69_{-0.07}^{+0.06}$ & $0.41_{-0.24}^{+0.66}$/$0.47_{-0.40}^{+0.60}$ & $1.93_{-0.18}^{+0.23}$/$-2.10_{-8.64}^{+2.13}$ & e \\ 
        
        f115d\_brt\_uds\_739 & 34.2446999 & -5.1153672 & 26.31 & 24.43 & $2.75_{-0.30}^{+0.21}$ & $2.96_{-0.08}^{+0.08}$ & $2.57_{-0.28}^{+0.38}$ & $9.46_{-0.20}^{+0.20}$/$10.01_{-0.01}^{+0.01}$ & .../$0.06_{-0.06}^{+0.15}$ & 0.8/$0.34_{-0.14}^{+0.10}$ & $0.03_{-0.02}^{+0.10}$/$1.07_{-0.54}^{+0.86}$ & $2.57_{-0.83}^{+0.91}$/$-22.33_{-14.78}^{+14.47}$ & c \\ 
        
        f115d\_brt\_uds\_813 & 34.3211879 & -5.2680748 & 26.42 & 22.18 & $2.26_{-0.18}^{+0.15}$ & $2.77_{-0.04}^{+0.04}$ & $2.04_{-0.28}^{+0.36}$ & $10.79_{-0.09}^{+0.10}$/$10.78_{-0.02}^{+0.01}$ & .../0 & 0.5/$0.61_{-0.18}^{+0.18}$ & $1.10_{-0.61}^{+0.66}$/$0.78_{-0.58}^{+0.86}$ & $1.63_{-0.44}^{+0.14}$/$-0.17_{-19.16}^{+0.48}$ & irr \\ 
        
        f150d\_brt\_uds\_046 & 34.4382923 & -5.2943129 & 25.16 & 23.55 & $2.81_{-0.16}^{+0.15}$ & $3.65_{-0.07}^{+0.07}$ & $2.68_{-0.28}^{+0.31}$ & $10.00_{-0.38}^{+0.29}$/$10.46_{-0.02}^{+0.01}$ & .../0 & 0.7/$0.50_{-0.10}^{+0.07}$ & $0.16_{-0.14}^{+0.46}$/$0.86_{-0.46}^{+0.56}$ & $1.86_{-0.30}^{+0.57}$/$-0.50_{-18.70}^{+0.27}$ & irr \\ 
        
        f150d\_brt\_uds\_069 & 34.3171538 & -5.2855252 & 26.44 & 25.17 & $2.84_{-0.16}^{+0.14}$ & $3.57_{-0.04}^{+0.04}$ & $2.97_{-0.24}^{+0.28}$ & $9.08_{-0.07}^{+0.11}$/$9.95_{-0.07}^{+0.01}$ & .../$0.02_{-0.02}^{+0.09}$ & 0.8/$0.28_{-0.09}^{+0.10}$ & $0.03_{-0.02}^{+0.03}$/$1.51_{-0.61}^{+0.56}$ & $1.63_{-0.19}^{+0.30}$/$-33.06_{-21.19}^{+21.61}$ & c \\ 
        
        f150d\_brt\_uds\_090 & 34.3865074 & -5.2751339 & 26.49 & 24.53 & $3.28_{-0.23}^{+0.20}$ & $3.88_{-0.01}^{+0.01}$ & $3.23_{-0.29}^{+0.30}$ & $9.81_{-0.09}^{+0.09}$/$10.48_{-0.03}^{+0.01}$ & .../0 & 0.8/$0.65_{-0.09}^{+0.09}$ & $0.04_{-0.01}^{+0.02}$/$0.67_{-0.51}^{+0.47}$ & $2.21_{-0.30}^{+0.15}$/$-0.34_{-14.54}^{+0.76}$ & d \\ 
        
        f150d\_brt\_uds\_143 & 34.404667 & -5.2547906 & 25.38 & 24.69 & $2.19_{-0.25}^{+0.20}$ & $2.24_{-0.07}^{+0.07}$ & $2.19_{-0.31}^{+0.40}$ & $9.47_{-0.51}^{+0.18}$/$9.70_{-0.02}^{+0.01}$ & .../$0.33_{-0.32}^{+0.32}$ & 1.0/$0.08_{-0.07}^{+0.11}$ & $2.76_{-0.70}^{+0.55}$/$2.24_{-0.83}^{+0.73}$ & $2.10_{-4.02}^{+0.18}$/$-48.63_{-31.30}^{+31.52}$ & c \\ 
        
        f150d\_brt\_uds\_181 & 34.375762 & -5.2418499 & 26.39 & 25.12 & $2.81_{-0.14}^{+0.14}$ & $3.04_{-0.03}^{+0.03}$ & $3.17_{-0.28}^{+0.32}$ & $9.53_{-0.25}^{+0.20}$/$10.14_{-0.03}^{+0.01}$ & .../$0.02_{-0.02}^{+0.06}$ & 0.7/$0.55_{-0.11}^{+0.10}$ & $0.08_{-0.06}^{+0.05}$/$0.73_{-0.56}^{+0.64}$ & $1.75_{-0.24}^{+0.34}$/$-0.08_{-16.02}^{+0.10}$ & c \\ 
        
        f150d\_brt\_uds\_334 & 34.3821957 & -5.1538846 & 26.07 & 24.88 & $3.19_{-0.20}^{+0.20}$ & $3.34_{-0.05}^{+0.05}$ & $2.92_{-0.34}^{+0.35}$ & $9.54_{-0.22}^{+0.31}$/$10.03_{-0.05}^{+0.03}$ & .../$0.08_{-0.08}^{+0.15}$ & 0.8/$0.43_{-0.16}^{+0.11}$ & $0.12_{-0.11}^{+0.15}$/$0.76_{-0.58}^{+0.74}$ & $2.01_{-0.58}^{+0.44}$/$-0.28_{-0.16}^{+0.09}$ & c \\ 
        
        f150d\_brt\_uds\_345 & 34.5014752 & -5.1511903 & 25.27 & 24.02 & $2.21_{-0.15}^{+0.15}$ & $2.91_{-0.05}^{+0.05}$ & $2.61_{-0.32}^{+0.36}$ & $9.51_{-0.35}^{+0.42}$/$10.26_{-0.03}^{+0.01}$ & .../$0.06_{-0.06}^{+0.14}$ & 0.7/$0.30_{-0.11}^{+0.11}$ & $0.23_{-0.21}^{+0.88}$/$1.71_{-0.79}^{+0.66}$ & $1.72_{-0.64}^{+1.44}$/$-36.93_{-23.64}^{+24.45}$ & c \\ 
        
        f150d\_brt\_uds\_346 & 34.4177096 & -5.1495225 & 25.47 & 23.80 & $3.24_{-0.20}^{+0.19}$ & $3.78_{-0.01}^{+0.01}$ & $3.11_{-0.26}^{+0.35}$ & $10.12_{-0.18}^{+0.34}$/$10.68_{-0.04}^{+0.01}$ & .../0 & 0.8/$0.66_{-0.10}^{+0.09}$ & $0.05_{-0.03}^{+0.19}$/$0.50_{-0.43}^{+0.62}$ & $2.36_{-0.30}^{+0.21}$/$0.17_{-8.13}^{+1.02}$ & d \\ 
        
        f150d\_brt\_uds\_427 & 34.3320032 & -5.1041668 & 25.83 & 24.49 & $2.77_{-0.18}^{+0.21}$ & $3.65_{-0.07}^{+0.07}$ & $2.97_{-0.36}^{+0.52}$ & $9.56_{-0.28}^{+0.43}$/$10.10_{-0.02}^{+0.01}$ & .../$0.06_{-0.06}^{+0.12}$ & 0.6/$0.36_{-0.13}^{+0.13}$ & $0.07_{-0.05}^{+0.94}$/$1.08_{-0.54}^{+0.73}$ & $1.91_{-0.69}^{+1.22}$/$-23.00_{-14.55}^{+14.95}$ & c \\ 
        \hline
    \end{tabular}
    }
    \tablecomments{
    Similar to Tables~\ref{tab:t1highz} and \ref{tab:t2highz} but for the Tier 1
    objects in the ``Low-$z$'' category. 
    {In CIGALE, the stellar masses are the values re-derived by rerunning it with the 
    three parameters above fixed at their 50th percentile}; this procedure is 
    different from that of the CIGALE run for the objects in the ``High-$z$'' 
    category, where the routine was run for only once and the stellar masses
    were obtained together with other parameters.
    The morphological type ``e''
    stands for ``elliptical''.
    }
\end{table*}

\begin{table*}[hbt!]
    \caption{Tier 2 ``Low-$z$'' Objects} \label{tab:t2lowz}
    \resizebox{\textwidth}{!}{
    \begin{tabular}{lccccccccccccc} \hline
       SID & R.A. & Decl. & $m_{\rm in}$ & $m_{356}$ & $z_{lp}$ & $z_{ez}$ & $z_{cg}$ & $\log(M_*)$ [$M_\odot$] & $f_{AGN}$ & $E(B-V)$ & Age (Gyr) & $\log$(SFR) [$M_\odot$/yr] & Morph \\ \hline 
       
        f115d\_brt\_ceers\_246 & 214.8402602 & 52.8011294 & 26.00 & 24.55 & $2.41_{-0.24}^{+0.25}$ & $2.32_{-0.06}^{+0.06}$ & $2.41_{-0.24}^{+0.24}$ & $9.00_{-0.07}^{+0.10}$/$9.80_{-0.04}^{+0.01}$ & .../0 & 0.9/$0.21_{-0.07}^{+0.07}$ & $0.01_{-0.00}^{+0.01}$/$1.70_{-0.80}^{+0.83}$ & $1.95_{-0.27}^{+0.19}$/$-13.99_{-37.18}^{+12.54}$ & c \\ 
        
        f115d\_brt\_ceers\_279 & 214.9417737 & 52.8845789 & 25.98 & 24.06 & $1.14_{-0.14}^{+0.17}$ & $2.64_{-0.06}^{+0.06}$ & $0.95_{-0.17}^{+0.22}$ & $9.66_{-0.15}^{+0.11}$/$9.52_{-0.02}^{+0.01}$ & .../$0.06_{-0.06}^{+0.12}$ & 0.6/$0.80_{-0.13}^{+0.17}$ & $3.82_{-1.30}^{+1.22}$/$3.91_{-1.02}^{+1.03}$ & $-0.67_{-0.24}^{+0.44}$/$-64.98_{-69.99}^{+63.25}$ & c \\ 
        
        f150d\_brt\_ceers\_078 & 214.7713801 & 52.7497666 & 26.31 & 24.49 & $3.26_{-0.26}^{+0.28}$ & $3.95_{-0.06}^{+0.06}$ & $1.30_{-0.27}^{+12.30}$ & $10.43_{-0.06}^{+0.05}$/- & .../- & 0.3/- & $1.75_{-0.29}^{+0.22}$/- & $2.52_{-2.07}^{+0.14}$/- & c \\

        f150d\_brt\_ceers\_165 & 214.7680323 & 52.8164203 & 26.15 & 23.37 & $2.80_{-0.14}^{+0.14}$ & $3.39_{-0.48}^{+0.37}$ & $3.32_{-0.46}^{+0.30}$ & $10.60_{-0.04}^{+0.04}$/$10.58_{-0.01}^{+0.01}$ & .../0 & 0.5/$0.17_{-0.08}^{+0.14}$ & $0.54_{-0.20}^{+0.07}$/$1.28_{-0.46}^{+0.40}$ & $0.64_{-2.16}^{+0.12}$/$-9.30_{-27.47}^{+8.79}$ & irr \\ 
        
        f090d\_brt\_cosmos\_013 & 150.0930808 & 2.1753794 & 24.79 & 22.46 & $1.68_{-0.18}^{+0.16}$ & $1.68_{-0.03}^{+0.03}$ & $1.16_{-0.17}^{+0.22}$ & $10.41_{-0.10}^{+0.06}$/$10.13_{-0.01}^{+0.01}$ & .../0 & 0.2/$0.30_{-0.05}^{+0.06}$ & $3.33_{-0.67}^{+0.51}$/$3.11_{-0.61}^{+0.39}$ & $-0.79_{-2.46}^{+4.53}$/$-23.69_{-72.91}^{+22.22}$ & d \\ 
        
        f090d\_brt\_cosmos\_051 & 150.0811983 & 2.1939242 & 24.62 & 21.91 & $1.97_{-0.27}^{+5.89}$ & $2.01_{-0.00}^{+0.00}$ & $1.24_{-0.15}^{+0.19}$ & -/$10.17_{-0.01}^{+0.01}$ & .../0 & 0.6/$0.62_{-0.04}^{+0.05}$ & -/$1.59_{-1.19}^{+2.02}$ & -/$0.86_{-0.01}^{+0.01}$ & d \\ 
        
        f090d\_brt\_cosmos\_099 & 150.0747367 & 2.2164983 & 24.56 & 21.76 & $7.00_{-0.14}^{+0.14}$ & $2.08_{-0.00}^{+0.00}$ & $1.82_{-0.38}^{+0.27}$ & -/$10.68_{-0.01}^{+0.01}$ & .../0 & -/$0.68_{-0.08}^{+0.10}$ & -/$0.60_{-0.53}^{+1.99}$ & -/$1.03_{-0.02}^{+0.02}$ & d \\ 
        
        f090d\_brt\_cosmos\_118 & 150.1026692 & 2.2248794 & 26.18 & 24.92 & $5.57_{-4.29}^{+0.18}$ & $1.20_{-0.15}^{+0.15}$ & $1.01_{-0.17}^{+0.19}$ & $9.12_{-0.05}^{+0.13}$/$8.88_{-0.01}^{+0.01}$ & .../0 & 0.4/$0.12_{-0.06}^{+0.06}$ & $0.10_{-0.02}^{+0.01}$/$3.19_{-1.35}^{+1.20}$ & $2.59_{-0.07}^{+0.07}$/$-25.72_{-71.80}^{+23.31}$ & c \\ 
        
        f090d\_brt\_cosmos\_122 & 150.1068407 & 2.227158 & 24.98 & 22.59 & $5.61_{-0.14}^{+0.14}$ & $1.81_{-0.02}^{+0.02}$ & $1.51_{-0.16}^{+0.19}$ & -/$10.24_{-0.13}^{+0.01}$ & .../0 & 0.6/$0.54_{-0.04}^{+0.05}$ & -/$3.43_{-1.00}^{+0.47}$ & -/$0.81_{-0.23}^{+0.01}$ & d \\ 
        
        f090d\_brt\_cosmos\_180 & 150.0905478 & 2.2441653 & 24.27 & 21.34 & $8.36_{-0.32}^{+0.21}$ & $1.99_{-0.02}^{+0.02}$ & $1.69_{-0.16}^{+0.21}$ & -/$10.76_{-0.01}^{+0.02}$ & .../0 & -/$0.46_{-0.04}^{+0.05}$ & -/$0.78_{-0.34}^{+0.28}$ & -/$0.26_{-13.61}^{+0.14}$ & d \\ 
        
        f090d\_brt\_cosmos\_507 & 150.0991287 & 2.3206993 & 25.43 & 22.52 & $8.36_{-0.49}^{+0.22}$ & $2.33_{-0.02}^{+0.02}$ & $2.07_{-0.21}^{+0.18}$ & -/$10.66_{-0.01}^{+0.01}$ & .../0 & -/$0.30_{-0.05}^{+0.06}$ & -/$2.43_{-1.61}^{+0.56}$ & -/$-18.98_{-53.49}^{+18.22}$ & d \\ 
        
        f090d\_brt\_cosmos\_611 & 150.1434149 & 2.3486737 & 25.38 & 23.86 & $1.51_{-0.19}^{+0.19}$ & $1.41_{-0.07}^{+0.07}$ & $1.04_{-0.19}^{+0.18}$ & $9.59_{-0.12}^{+0.09}$/$8.98_{-0.01}^{+0.01}$ & .../0 & 0.1/$0.43_{-0.13}^{+0.04}$ & $2.02_{-0.76}^{+0.75}$/$0.32_{-0.30}^{+1.97}$ & $-0.61_{-0.59}^{+0.27}$/$-0.47_{-0.02}^{+0.02}$ & c \\ 
        
        f090d\_brt\_cosmos\_825 & 150.1733887 & 2.4007302 & 24.87 & 22.70 & $2.93_{-0.20}^{+0.23}$ & $3.59_{-0.02}^{+0.02}$ & $3.16_{-0.27}^{+0.55}$ & $9.97_{-0.06}^{+0.14}$/$10.64_{-0.01}^{+0.01}$ & .../0 & 0.6/$0.35_{-0.06}^{+0.05}$ & $0.03_{-0.01}^{+0.01}$/$1.44_{-1.27}^{+0.61}$ & $2.55_{-0.13}^{+0.55}$/$1.74_{-0.01}^{+0.01}$ & irr \\ 
        
        f090d\_brt\_cosmos\_882 & 150.1479556 & 2.4300038 & 25.21 & 22.14 & $1.58_{-0.14}^{+0.16}$ & $1.87_{-0.01}^{+0.01}$ & $1.22_{-0.15}^{+0.20}$ & -/$10.16_{-0.01}^{+0.01}$ & .../0 & 0.7/$0.62_{-0.04}^{+0.04}$ & -/$3.19_{-0.37}^{+0.33}$ & -/$0.58_{-0.01}^{+0.01}$ & d \\ 
        
        f090d\_brt\_cosmos\_939 & 150.1830083 & 2.4695256 & 24.09 & 22.50 & $1.58_{-0.14}^{+0.15}$ & $1.65_{-0.04}^{+0.04}$ & $1.23_{-0.15}^{+0.19}$ & -/$9.99_{-0.01}^{+0.01}$ & .../0 & 0.4/$0.37_{-0.04}^{+0.04}$ & -/$3.40_{-0.51}^{+0.98}$ & -/$0.35_{-0.01}^{+0.01}$ & d \\ 
        
        f115d\_brt\_cosmos\_009 & 150.1010045 & 2.1736321 & 24.29 & 21.76 & $6.39_{-0.15}^{+0.16}$ & $2.41_{-0.00}^{+0.00}$ & $1.74_{-0.22}^{+0.24}$ & -/$10.72_{-0.01}^{+0.01}$ & .../0 & -/$0.82_{-0.09}^{+0.12}$ & -/$0.62_{-0.56}^{+2.48}$ & -/$0.86_{-0.04}^{+0.03}$ & d \\ 
        
        f115d\_brt\_cosmos\_073 & 150.0969656 & 2.2288968 & 24.55 & 22.32 & $7.79_{-0.14}^{+0.14}$ & $2.48_{-0.02}^{+0.02}$ & $1.98_{-0.15}^{+0.19}$ & -/$10.33_{-0.01}^{+0.01}$ & .../0 & -/$0.70_{-0.04}^{+0.04}$ & -/$0.77_{-0.40}^{+0.36}$ & -/$1.66_{-0.00}^{+0.00}$ & d \\ 
        
        f115d\_brt\_cosmos\_142 & 150.0603007 & 2.2673313 & 24.67 & 22.74 & $8.76_{-0.15}^{+0.16}$ & $1.97_{-0.02}^{+0.02}$ & $1.52_{-0.17}^{+0.19}$ & -/$9.95_{-0.01}^{+0.01}$ & .../0 & -/$0.60_{-0.09}^{+0.04}$ & -/$0.33_{-0.31}^{+2.29}$ & -/$-2.71_{-4.75}^{+1.73}$ & d \\ 
        
        f115d\_brt\_cosmos\_208 & 150.057066 & 2.292873 & 24.56 & 21.76 & $4.00_{-0.14}^{+0.14}$ & $3.13_{-0.03}^{+0.03}$ & $1.99_{-0.15}^{+0.20}$ & $11.41_{-0.04}^{+0.03}$/$10.92_{-0.01}^{+0.01}$ & .../0 & 0.4/$0.69_{-0.04}^{+0.04}$ & $0.37_{-0.03}^{+0.03}$/$0.83_{-0.29}^{+0.28}$ & $-4.16_{-1.31}^{+1.66}$/$-5.95_{-16.30}^{+5.68}$ & irr \\ 
        
        f115d\_brt\_cosmos\_338 & 150.1748959 & 2.3527429 & 25.21 & 23.20 & $9.15_{-0.45}^{+1.34}$ & $3.08_{-0.03}^{+0.03}$ & $2.96_{-0.22}^{+0.31}$ & -/$10.39_{-0.01}^{+0.09}$ & .../0 & -/$0.57_{-0.06}^{+0.06}$ & -/$0.78_{-0.52}^{+0.49}$ & -/$1.92_{-0.05}^{+0.01}$ & irr \\ 
        
        f150d\_brt\_cosmos\_119 & 150.0597858 & 2.2810155 & 25.22 & 23.35 & $3.42_{-1.29}^{+0.30}$ & $4.23_{-0.08}^{+0.08}$ & $2.09_{-0.21}^{+0.18}$ & $10.20_{-0.04}^{+0.04}$/$10.52_{-0.01}^{+0.01}$ & .../0 & 1.0/$0.68_{-0.06}^{+0.04}$ & $1.49_{-0.43}^{+0.20}$/$1.79_{-0.65}^{+0.58}$ & $3.48_{-0.08}^{+0.09}$/$-13.08_{-39.87}^{+12.30}$ & d \\ 
        
        f150d\_brt\_cosmos\_182 & 150.1915483 & 2.3146803 & 26.45 & 25.17 & $2.24_{-0.19}^{+0.85}$ & $3.39_{-0.15}^{+0.15}$ & $3.01_{-0.37}^{+0.44}$ & $9.41_{-0.29}^{+0.26}$/$9.84_{-0.01}^{+0.01}$ & .../0 & 0.9/$0.33_{-0.09}^{+0.15}$ & $0.04_{-0.02}^{+1.27}$/$1.22_{-0.57}^{+0.69}$ & $2.68_{-0.95}^{+0.97}$/$-9.72_{-26.42}^{+8.72}$ & c \\ 
        
        f150d\_brt\_cosmos\_257 & 150.1625839 & 2.3540574 & 24.95 & 23.75 & $8.74_{-0.14}^{+0.17}$ & $2.17_{-0.09}^{+0.09}$ & $1.79_{-0.19}^{+0.19}$ & -/$9.95_{-0.02}^{+0.01}$ & .../0 & -/$0.60_{-0.06}^{+0.04}$ & -/$0.85_{-0.29}^{+0.31}$ & -/$-6.50_{-17.24}^{+5.42}$ & d \\ 

        
        f200d\_brt\_cosmos\_389 & 150.1006761 & 2.3348299 & 25.81 & 24.93 & $4.59_{-0.40}^{+0.21}$ & $5.72_{-0.04}^{+0.04}$ & $4.95_{-0.50}^{+0.42}$ & $10.49_{-0.65}^{+0.10}$/$10.67_{-0.01}^{+0.01}$ & .../0 & 0.7/$0.54_{-0.08}^{+0.08}$ & $0.59_{-0.55}^{+0.35}$/$0.48_{-0.40}^{+0.44}$ & $1.79_{-0.15}^{+0.84}$/$-2.94_{-8.70}^{+3.00}$ & d \\ 

        f090d\_brt\_jsmed\_157 & 53.0790797 & -27.8730445 & 27.07 & 25.05 & $0.10_{-0.07}^{+0.07}$ & $0.89_{-0.01}^{+0.01}$ & $0.74_{-0.16}^{+0.19}$ & -/$8.30_{-0.01}^{+0.01}$ & .../0 & -/$0.04_{-0.04}^{+0.06}$ & -/$5.97_{-1.48}^{+0.36}$ & -/$-48.82_{-140.11}^{+45.36}$ & d \\ 

        f090d\_brt\_jsmed\_189 & 53.0475035 & -27.8704957 & 26.43 & 22.39 & $4.16_{-0.14}^{+0.16}$ & $3.72_{-0.01}^{+0.01}$ & $3.98_{-0.14}^{+0.21}$ & $11.11_{-0.04}^{+0.03}$/$10.65_{-0.01}^{+0.01}$ & .../0 & 0.2/$0.29_{-0.04}^{+0.05}$ & $0.49_{-0.04}^{+0.05}$/$0.22_{-0.21}^{+0.29}$ & $-0.43_{-0.14}^{+0.26}$/$-1.20_{-2.10}^{+0.83}$ & irr \\ 
        
        f090d\_brt\_jsmed\_386 & 53.0875763 & -27.840385 & 27.09 & 23.79 & $1.39_{-0.14}^{+0.14}$ & $2.02_{-0.00}^{+0.00}$ & $1.58_{-0.23}^{+0.25}$ & $9.17_{-0.40}^{+0.30}$/$9.88_{-0.01}^{+0.01}$ & .../0 & 0.8/$0.29_{-0.06}^{+0.07}$ & $0.39_{-0.36}^{+0.34}$/$3.11_{-0.64}^{+0.52}$ & $1.10_{-0.46}^{+1.63}$/$-24.57_{-71.51}^{+22.91}$ & irr \\ 
        
        f090d\_brt\_uds\_014 & 34.4826496 & -5.2862261 & 25.65 & 23.00 & $1.84_{-0.18}^{+0.15}$ & $1.75_{-0.06}^{+0.06}$ & $1.23_{-0.20}^{+0.24}$ & $10.14_{-0.16}^{+0.14}$/$9.88_{-0.03}^{+0.01}$ & .../$0.35_{-0.31}^{+0.31}$ & 0.2/$0.37_{-0.16}^{+0.10}$ & $1.57_{-0.51}^{+1.09}$/$1.72_{-0.94}^{+1.71}$ & $-0.12_{-0.68}^{+3.72}$/$-37.08_{-24.09}^{+24.23}$ & c \\ 

        f090d\_brt\_uds\_055 & 34.4650194 & -5.2724489 & 25.61 & 25.48 & $0.60_{-0.14}^{+0.14}$ & $0.95_{-0.01}^{+0.01}$ & $1.35_{-0.32}^{+0.30}$ & $7.69_{-0.40}^{+0.26}$/$8.65_{-0.01}^{+0.01}$ & .../$0.57_{-0.46}^{+0.48}$ & 1.0/$0.03_{-0.03}^{+0.04}$ & $0.23_{-0.22}^{+2.21}$/$0.57_{-0.46}^{+0.48}$ & $0.87_{-0.41}^{+0.28}$/$-12.48_{-6.74}^{+6.94}$ & c \\ 

        f090d\_brt\_uds\_107 & 34.495778 & -5.2554256 & 25.57 & 22.60 & $8.59_{-0.15}^{+0.16}$ & $2.37_{-0.07}^{+0.07}$ & $1.98_{-0.17}^{+0.21}$ & -/$10.55_{-0.01}^{+0.01}$ & 0 & -/$0.54_{-0.05}^{+0.06}$ & -/$1.05_{-0.93}^{+1.77}$ & -/$1.23_{-0.01}^{+0.01}$ & d \\ 
        
        f090d\_brt\_uds\_276 & 34.3264188 & -5.1373572 & 26.69 & 23.68 & $5.40_{-0.14}^{+0.14}$ & $2.91_{-0.01}^{+0.01}$ & $5.94_{-0.17}^{+0.20}$ & $10.25_{-0.03}^{+0.04}$/$10.64_{-0.01}^{+0.01}$ & .../0 & 0.6/$0.12_{-0.04}^{+0.04}$ & $0.11_{-0.01}^{+0.02}$/$0.22_{-0.21}^{+0.27}$ & $3.37_{-0.08}^{+0.09}$/$-0.88_{-2.07}^{+0.83}$ & irr \\ 

        f090d\_brt\_uds\_292 & 34.3199395 & -5.1346201 & 24.93 & 21.81 & $1.68_{-0.17}^{+0.16}$ & $2.14_{-0.00}^{+0.00}$ & $1.16_{-0.21}^{+0.23}$ & -/$10.26_{-0.01}^{+0.01}$ & .../$0.07_{-0.07}^{+0.15}$ & -/$0.78_{-0.08}^{+0.08}$ & -/$0.76_{-0.59}^{+2.12}$ & -/$-14.28_{-11.87}^{+11.60}$ & c \\ 
        
        f090d\_brt\_uds\_322 & 34.419778 & -5.1265324 & 24.54 & 22.90 & $1.60_{-0.14}^{+0.14}$ & $1.47_{-0.03}^{+0.03}$ & $1.43_{-0.24}^{+0.23}$ & $10.90_{-0.17}^{+0.05}$/$10.03_{-0.01}^{+0.01}$ & .../$0.41_{-0.36}^{+0.36}$ & 0.0/$0.04_{-0.04}^{+0.05}$ & $3.56_{-0.47}^{+0.74}$/$3.09_{-0.97}^{+0.88}$ & $2.94_{-6.73}^{+0.22}$/$-66.47_{-43.08}^{+43.24}$ & c \\ 
        
        f090d\_brt\_uds\_324 & 34.3416163 & -5.1254591 & 24.27 & 21.88 & $1.59_{-0.14}^{+0.14}$ & $1.72_{-0.00}^{+0.00}$ & $1.05_{-0.19}^{+0.18}$ & $10.87_{-0.12}^{+0.07}$/$10.07_{-0.01}^{+0.01}$ & .../0 & 0.2/$0.55_{-0.06}^{+0.07}$ & -/$0.82_{-0.41}^{+0.59}$ & -/$-5.92_{-16.68}^{+5.05}$ & d \\ 
        
        f115d\_brt\_uds\_161 & 34.3434954 & -5.2761198 & 24.48 & 22.91 & $1.60_{-0.14}^{+0.14}$ & $2.47_{-0.02}^{+0.02}$ & $2.34_{-0.26}^{+0.23}$ & $10.26_{-0.06}^{+0.08}$/$10.49_{-0.02}^{+0.01}$ & .../0 & 0.5/$0.08_{-0.06}^{+0.06}$ & $0.68_{-0.14}^{+0.17}$/$2.25_{-0.69}^{+0.51}$ & $2.04_{-0.73}^{+1.30}$/$-17.47_{-50.14}^{+16.51}$ & irr \\ 
        
        f115d\_brt\_uds\_193 & 34.403946 & -5.268707 & 26.72 & 23.95 & $2.07_{-0.18}^{+0.17}$ & $3.09_{-0.09}^{+0.09}$ & $2.86_{-0.38}^{+0.39}$ & $11.27_{-0.32}^{+0.05}$/$10.39_{-0.03}^{+0.01}$ & .../$0.08_{-0.08}^{+0.16}$ & 1.0/$0.28_{-0.12}^{+0.12}$ & $0.02_{-0.01}^{+1.01}$/$1.68_{-0.66}^{+0.58}$ & $2.14_{-0.77}^{+0.48}$/$-35.50_{-24.46}^{+23.48}$ & c \\ 
        
        f115d\_brt\_uds\_221 & 34.34927 & -5.2634548 & 24.11 & 21.73 & $6.94_{-0.14}^{+0.17}$ & $2.40_{-0.08}^{+0.08}$ & $1.70_{-0.27}^{+0.28}$ & -/$10.74_{-0.02}^{+0.01}$ & .../$0.15_{-0.15}^{+0.19}$ & -/$0.68_{-0.10}^{+0.08}$ & -/$0.79_{-0.54}^{+0.74}$ & -/$-5.25_{-18.69}^{+5.55}$ & irr \\ 
        
        f115d\_brt\_uds\_263 & 34.2904198 & -5.2531251 & 27.05 & 24.88 & $2.06_{-0.23}^{+0.41}$ & $3.03_{-0.11}^{+0.11}$ & $2.29_{-0.34}^{+0.45}$ & $9.69_{-0.57}^{+0.72}$/$9.86_{-0.02}^{+0.01}$ & .../0 & 0.9/$0.36_{-0.08}^{+0.07}$ & $0.03_{-0.02}^{+0.08}$/$1.91_{-0.94}^{+0.92}$ & $1.34_{-0.32}^{+0.26}$/$-15.17_{-42.63}^{+13.78}$ & d \\ 
        
        f115d\_brt\_uds\_606 & 34.4056603 & -5.1556666 & 24.80 & 23.25 & $2.20_{-0.14}^{+0.14}$ & $2.46_{-0.00}^{+0.00}$ & $2.14_{-0.24}^{+0.29}$ & $9.57_{-0.21}^{+0.24}$/$10.31_{-0.02}^{+0.01}$ & .../$0.20_{-0.20}^{+0.28}$ & 0.2/$0.11_{-0.07}^{+0.07}$ & $3.16_{-0.24}^{+0.26}$/$2.56_{-0.69}^{+0.60}$ & $-3.45_{-0.77}^{+7.38}$/$-55.14_{-35.41}^{+37.26}$ & c \\
        
        f115d\_brt\_uds\_620 & 34.3343455 & -5.1520742 & 24.89 & 23.47 & $2.40_{-0.14}^{+0.14}$ & $2.57_{-0.00}^{+0.00}$ & $2.28_{-0.21}^{+0.22}$ & $11.00_{-0.03}^{+0.03}$/$10.18_{-0.02}^{+0.01}$ & .../0 & 0.4/$0.36_{-0.06}^{+0.05}$ & $0.62_{-0.05}^{+0.06}$/$0.74_{-0.50}^{+0.40}$ & $-3.09_{-0.59}^{+6.92}$/$-5.41_{-14.32}^{+4.67}$ & d \\ 
        
        f115d\_brt\_uds\_666 & 34.3843112 & -5.140298 & 26.63 & 23.80 & $8.68_{-0.18}^{+0.16}$ & $2.92_{-0.03}^{+0.03}$ & $2.20_{-0.29}^{+0.34}$ & -/$10.25_{-0.01}^{+0.01}$ & .../$0.24_{-0.23}^{+0.23}$ & -/$0.68_{-0.17}^{+0.12}$ & -/$1.09_{-0.72}^{+1.24}$ & -/$-22.90_{-15.19}^{+15.31}$ & c \\ 
        
        f115d\_brt\_uds\_667 & 34.2848298 & -5.1407918 & 24.49 & 22.03 & $2.13_{-0.16}^{+0.17}$ & $2.22_{-0.20}^{+0.20}$ & $1.98_{-0.17}^{+0.23}$ & $9.34_{-0.31}^{+0.58}$/$10.89_{-0.03}^{+0.01}$ & .../$0.12_{-0.12}^{+0.17}$ & 0.4/$0.41_{-0.10}^{+0.06}$ & $1.65_{-0.37}^{+0.92}$/$2.47_{-0.76}^{+0.71}$ & $0.71_{-0.10}^{+0.38}$/$-53.15_{-33.41}^{+35.53}$ & c \\ 
        
        f115d\_brt\_uds\_755 & 34.2664134 & -5.1089223 & 24.31 & 22.98 & $2.14_{-0.26}^{+0.34}$ & $2.49_{-0.01}^{+0.01}$ & $2.44_{-0.19}^{+0.23}$ & $11.11_{-0.03}^{+0.03}$/$10.45_{-0.02}^{+0.01}$ & .../$0.47_{-0.35}^{+0.35}$ & 0.4/$0.03_{-0.03}^{+0.04}$ & $0.28_{-0.05}^{+0.35}$/$2.06_{-0.61}^{+0.56}$ & $0.38_{-0.40}^{+0.56}$/$-43.72_{-29.02}^{+28.93}$ & c \\ 
        
        f150d\_brt\_uds\_063 & 34.3321593 & -5.2880617 & 25.34 & 23.67 & $11.18_{-0.30}^{+0.34}$ & $3.43_{-0.03}^{+0.03}$ & $3.62_{-0.38}^{+0.32}$ & -/$10.57_{-0.01}^{+0.01}$ & .../0 & -/$0.46_{-0.06}^{+0.07}$ & -/$0.25_{-0.24}^{+0.38}$ & -/$0.28_{-3.25}^{+1.11}$ & d \\ 
        
        f150d\_brt\_uds\_088 & 34.2522909 & -5.275846 & 26.50 & 25.50 & $2.08_{-0.27}^{+0.52}$ & $2.81_{-0.19}^{+0.19}$ & $2.85_{-0.69}^{+0.54}$ & $9.57_{-0.21}^{+0.24}$/$9.60_{-0.04}^{+0.01}$ & .../$0.31_{-0.31}^{+0.31}$ & 1.0/$0.11_{-0.09}^{+0.17}$ & $1.34_{-1.33}^{+0.89}$/$1.60_{-0.75}^{+0.71}$ & $1.34_{-1.31}^{+0.75}$/$-34.16_{-21.63}^{+21.89}$ & c \\ 
        
        f150d\_brt\_uds\_158 & 34.3031646 & -5.2504972 & 26.37 & 25.49 & $3.39_{-0.41}^{+7.41}$ & $3.44_{-0.16}^{+0.16}$ & $2.17_{-0.52}^{+0.87}$ & $11.00_{-0.03}^{+0.03}$/$9.45_{-0.07}^{+0.01}$ & .../0 & 0.4/$0.29_{-0.15}^{+0.13}$ & $0.03_{-0.01}^{+0.11}$/$1.52_{-0.91}^{+1.45}$ & $2.62_{-0.97}^{+0.75}$/$-11.41_{-32.89}^{+9.87}$ & c \\ 
        
        f150d\_brt\_uds\_159 & 34.4929619 & -5.2500532 & 25.57 & 24.26 & $2.80_{-0.14}^{+0.14}$ & $3.53_{-0.03}^{+0.03}$ & $2.82_{-0.29}^{+0.33}$ & $9.34_{-0.31}^{+0.58}$/$10.26_{-0.05}^{+0.01}$ & .../$0.03_{-0.03}^{+0.10}$ & 0.7/$0.42_{-0.15}^{+0.10}$ & $0.03_{-0.02}^{+0.08}$/$0.92_{-0.49}^{+0.68}$ & $2.57_{-0.77}^{+1.17}$/$-10.62_{-18.68}^{+10.00}$ & c \\ 
        
        f150d\_brt\_uds\_333 & 34.2934613 & -5.1547235 & 24.73 & 23.02 & $8.80_{-0.14}^{+0.14}$ & $2.62_{-0.03}^{+0.03}$ & $2.06_{-0.23}^{+0.21}$ & -/$10.38_{-0.01}^{+0.01}$ & .../0 & -/$0.83_{-0.07}^{+0.08}$ & -/$1.43_{-1.09}^{+1.43}$ & -/$1.21_{-0.01}^{+0.01}$ & c \\ 
        
        f200d\_brt\_uds\_133 & 34.2840954 & -5.2605105 & 25.23 & 24.52 & $2.79_{-0.14}^{+0.14}$ & $5.06_{-0.08}^{+0.08}$ & $2.86_{-0.26}^{+0.22}$ & $9.69_{-0.57}^{+0.72}$/$10.43_{-0.01}^{+0.01}$ & .../0 & 0.9/$0.58_{-0.09}^{+0.09}$ & $0.05_{-0.02}^{+0.07}$/$1.92_{-0.48}^{+0.36}$ & $1.94_{-0.23}^{+0.15}$/$-14.39_{-42.56}^{+13.66}$ & d \\ 
        
        f200d\_brt\_uds\_154 & 34.2734842 & -5.2513303 & 26.23 & 25.36 & $2.61_{-0.20}^{+0.22}$ & $5.86_{-0.17}^{+0.17}$ & $2.62_{-0.52}^{+0.54}$ & $9.57_{-0.21}^{+0.24}$/$10.03_{-0.05}^{+0.01}$ & .../0 & 0.9/$0.77_{-0.18}^{+0.16}$ & $0.16_{-0.13}^{+1.72}$/$1.54_{-0.86}^{+0.98}$ & $1.33_{-0.63}^{+0.95}$/$0.48_{-24.93}^{+0.31}$ & c \\ 
        \hline 
    \end{tabular}
    }
    \tablecomments{
    Similar to Table \ref{tab:t1lowz} but for the Tier 2 objects in the 
    ``Low-$z$'' category. 
    }
\end{table*}

$\bullet$ {\bf Undecided: } This category contains 34 objects that cannot be 
placed in either category above. For the sake of completeness, they are listed 
in Table~\ref{tab:undecided}.

    In the $m_{115}-m_{356}$ versus $m_{356}$ distribution shown in 
Figure~\ref{fig:115-356_color}, the objects in the above three categories 
are indicated by different colors. It is obvious that many ``High-$z$'' objects
can be classified as EROs under the fiducial criterion of 
$m_{115}-m_{356}>2.0$~mag. In total, 80\%, 87\% and 63\% of the objects in the 
``High-$z$'', ``Low-$z$'' and ``Undecided'' categories can be classified as EROs, 
respectively.

\begin{figure*}
    \centering
    \includegraphics[width=0.95\linewidth]{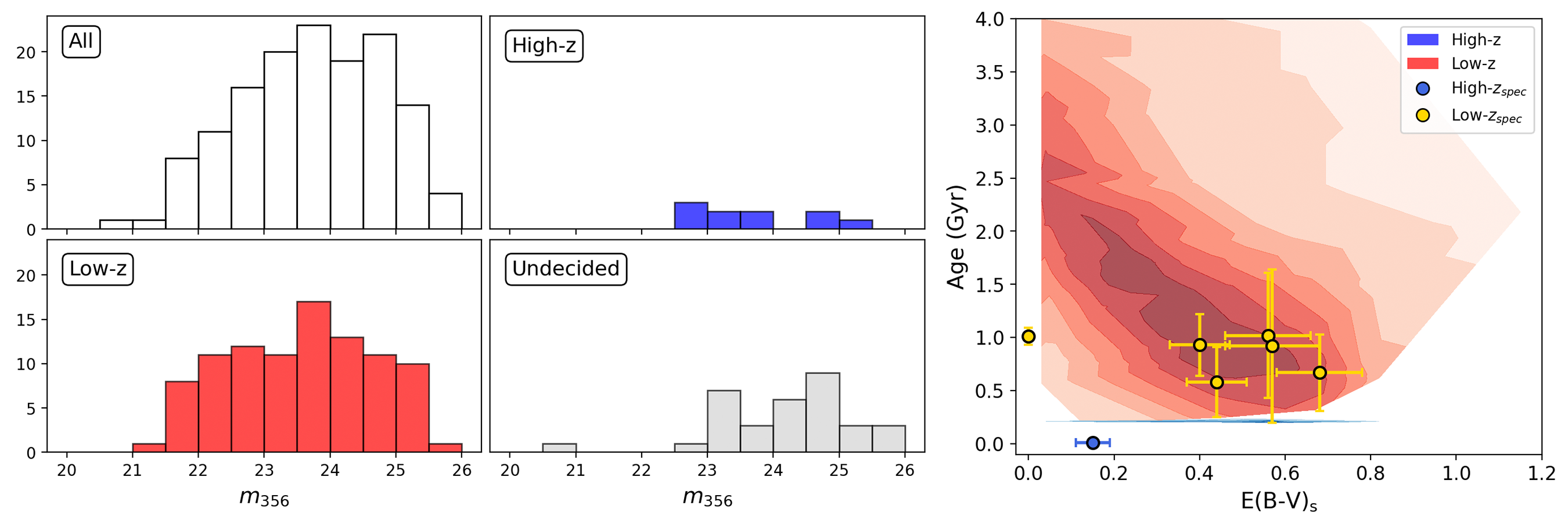}
    \caption{
    {\it Left: } F356W magnitude distributions for all objects, and objects in 
    the ``High-$z$'', ``Low-$z$'', and ``Undecided'' categories, respectively. 
    {\it Right: } Contour plots of Age vs. E(B-V) for the objects in the 
    ``High-$z$'' 
    (blue) and ``Low-$z$'' (red) categories. Both parameters are the 50th
    percentile values from the CIGALE run (using those from the Le Phare run
    gives similar results).
    The overlaid data points represent the SED fitting results of the seven 
    spectroscopically confirmed 
    grade I objects in Table~\ref{tab:spec-z} (see Section 6) when fixing the
    redshifts to their $z_{\rm spec}$;
    the only confirmed high-$z$ object is shown in blue, while all other objects 
    (all at low redshifts) are shown in gold. 
    }
    \label{fig:age_ebv_all}
\end{figure*}

    Figure~\ref{fig:age_ebv_all} shows the distribution of the fitted age and 
$E(B-V)$ for the ``Low-$z$'' and ``High-$z$'' objects, which are the two most
important (and degenerated) parameters that might give low-$z$ galaxies red colors
mimicking those of high-$z$ galaxies. As evident in the figure, the regions
occupied by the two categories are not well separated. The contours of the 
``High-$z$'' objects make a thin slab at the age of $\sim$0.25~Gyr, which is
largely due to the constraint that the age of a galaxy cannot be older than that
of the universe at the fitted high redshift. However, the region of the 
``Low-$z$'' objects also extends to very close to this area. 

{Using the SED fitting results by CIGALE, 
we also examined the fraction of stellar mass formed in the recent burst 
for each object in the ``High-$z$" and ``Low-$z$" categories. 
We found that there are significant differences: 
the ``High-$z$" candidates tend to have large burst fractions (median $\sim$55\%), 
suggesting that a considerable portion of their stellar mass had been formed in a recent burst. 
In contrast, the ``Low-$z$" candidates exhibit little burst activity (median near zero). 
Considering the generally young ages of the ``High-$z$" candidates, 
these results further suggests that they are dominated by young and bursty star-forming populations, 
whereas the ``Low-$z$" objects are characterized by older stellar populations with non-bursty star formation histories. 
}
    
    Another interesting point revealed in our SED analysis is that incorporating 
mid-IR photometry, while being helpful, still does not provide a decisive factor 
to separate the low-$z$ and high-$z$ candidates. For example, contrary to what one might
naively believe, being bright in mid-IR does not necessarily preclude a good SED
fit to give a high-$z$ solution (see Figure~\ref{fig:sed_example}).

\section{Spectroscopic Identifications}

    As it turns out, ten of our bright dropouts in the main sample have
existing NIRSpec spectroscopic data, which we used to identify their 
redshifts. These data are from the following programs: 
PID 1213 (PI N. Luetzgendorf), 
1215 (PI N. Luetzgendorf), 
4233 (PIs A. de Graaff \& G. Brammer), and 
6585 (PI D. Coulter). 
We used the data taken in the PRISM mode, which 
cover the range of 0.6--5.3~$\mu$m with the resolving power of 
$R\approx 30$--300.

    We reduced these data on our own. We first retrieved the Level 1b products 
from MAST and processed them through the {\tt calwebb\_detector1} step of the 
JWST pipeline (version 1.14.0) in the context of {\tt jwst\_1234.pmap}. The 
output ``rate.fits'' files were then processed through the {\sc msaexp} 
package \citep[][version 0.8.4]{Brammar23_msaexp}, which provides an 
end-to-end reduction including the final extraction of spectra. Briefly, the 
procedure removes the so-called ``$1/f$'' noise pattern, detects and masks the 
``snowball'' defects, subtracts the bias level, applies the flat-field, 
corrects the path-loss, does the flux calibration, traces spectra on all 
single exposures, and combines the single exposures with outlier rejection. 
All data used in this work were taken under the 3-shutter setup. In most
cases, the background was subtracted using the measurements in the nearest 
blank slit before drizzling individual exposures onto a common pixel grid. A 
few of our sources extend to all shutters, and the background could not be 
estimated locally. In such cases, the background was estimated using a nearby 
slit that was relatively blank. 

   Among the ten objects, seven have highly reliable redshifts based on at 
least two emission lines, which we rank as grade ``\Romannum{1}''. Two have
only a single line, which we assume to be H$\alpha$ due to a marginal 
detection that could be [O~\textsc{\romannum{3}}] that yields a consistent 
redshift. These two identifications are ranked as grade ``\Romannum{2}''. One
other does not show any emission 
lines and cannot be identified, which we assign grade ``\Romannum{3}''. 
The results are summarized in Table~\ref{tab:spec-z}. 

   While still limited, these spectroscopic identifications are broadly 
consistent with our SED analysis that most of the bright dropouts should be at 
low redshifts.
In terms of the accuracy of our categorization, the picture is mixed.  
Among the seven with grade \Romannum{1} $z_{\rm spec}$, two are the candidates in 
our ``Low-$z$'' category, three are in the ``High-$z$'' category, and two 
are in the ``Undecided'' category. The two ``Low-$z$'' ones are indeed confirmed
to be at $z\approx 3$. One of the three ``High-$z$'' candidates, which is a T2 
F115W dropout with a compact morphology, is confirmed at $z_{\rm spec}=8.679$ (this 
is a recovery of the identification by \citet[][]{Zitrin2015b} made in the 
pre-JWST era, which will be further discussed in Section 7). However, two other 
``High-$z$'' candidates  (one T1 F150W dropout and one T2 F115W dropout) turn out 
to be at low redshifts. The two ``Undecided'' objects are at low redshifts as 
well. In other words, our ``Low-$z$'' category seems to be robust, but the 
``High-$z$'' category is severely contaminated. There are three T1 and seven T2
``High-$z$'' objects in our main sample, and one in each tier has been refuted. 
Nonetheless, the one that is confirmed at $z_{\rm spec}=8.679$ shows that there 
indeed could be bona fide high-$z$ objects in our ``High-$z$'' category.

\begin{table*}[hbt!]
    \centering
    \caption{NIRSpec/MSA Identifications of Very Bright Dropouts}
    \resizebox{\textwidth}{!}{
    \begin{tabular}{lccccc||ccc} \hline
        SID & Category (Tier)/Morph & $z_{\rm spec}$ & $z_{\rm lp}$ & $z_{\rm ez}$ & $z_{\rm cg}$ & T (Myr) & E(B-V) & $\log_{10}(M_*/M_\odot)$ \\ \hline
        f115d\_brt\_uds\_245 & High-z (T2)/irr & 2.53 (\Romannum{1}) & $8.78_{-0.14}^{+0.15}$ & $2.46_{-0.02}^{+0.02}$ & $8.55_{-1.47}^{+1.47}$ & -/$576\pm328$ & -/$0.44\pm0.07$ & -/$10.61\pm0.06$ \\ 
        f115d\_brt\_cosmos\_344 & Undecided/irr & 2.99 (\Romannum{1}) & $8.49_{-0.49}^{+0.24}$ & $3.10_{-0.03}^{+0.03}$ & $3.38_{-1.26}^{+1.26}$ & -/$1015\pm592$ & -/$0.56\pm0.10$ & -/$10.69\pm0.14$ \\ 
        f150d\_brt\_ceers\_113 & Undecided/c & 3.10 (\Romannum{1}) & $15.38_{-0.14}^{+0.15}$ & $3.98_{-0.01}^{+0.01}$ & $3.09_{-2.18}^{+2.18}$ & -/$934\pm288$ & -/$0.40\pm0.07$ & -/$10.40\pm0.05$ \\ 
        f115d\_brt\_ceers\_279 & Low-z (T2)/c & 3.21 (\Romannum{1}) & $1.14_{-0.14}^{+0.17}$ & $2.64_{-0.06}^{+0.06}$ & $1.01_{-0.22}^{+0.22}$ & $1575_{-524}^{+217}$/$1010\pm87$ & 0.0/$0.00\pm0.01$ & $10.08_{-0.25}^{+0.05}$/$9.99\pm0.02$ \\ 
        f115d\_brt\_uds\_685 & Low-z (T1)/e & 3.23 (\Romannum{1}) & $3.17_{-0.16}^{+0.17}$ & $3.41_{-0.06}^{+0.06}$ & $2.74_{-0.24}^{+0.24}$ & $229_{-206}^{+710}$/$913\pm718$ & 1.0/$0.57\pm0.11$ & $10.25_{-0.25}^{+0.44}$/$10.77\pm0.11$ \\ 
        f150d\_brt\_ceers\_051 & High-z (T1)/c & 3.46 (\Romannum{1}) & $12.22_{-0.28}^{+0.24}$ & $14.57_{-0.01}^{+0.01}$ & $14.37_{-0.80}^{+0.80}$ & $34_{-19}^{+71}$/$665\pm360$ & 0.8/$0.68\pm0.10$ & $9.84_{-0.22}^{+0.18}$/$10.38\pm0.10$ \\ 
        f115d\_brt\_ceers\_062 & High-z (T2)/c & 8.679 (\Romannum{1}) & $8.80_{-0.14}^{+0.14}$ & $8.88_{-0.01}^{+0.01}$ & $8.95_{-0.12}^{+0.12}$ & -/$13\pm8$ & -/$0.15\pm0.04$ & -/$9.10\pm0.08$ \\ 
        \hline
        f150d\_brt\_uds\_158 & Low-z (T2)/c & 3.72 (\Romannum{2}) & $3.39_{-0.41}^{+7.41}$ & $3.44_{-0.16}^{+0.16}$ & $2.45_{-0.90}^{+0.90}$ & $82_{-63}^{+454}$/$1073\pm365$ & 0.4/$0.08\pm0.10$ & $9.22_{-0.25}^{+0.38}$/$9.75\pm0.07$ \\ 
        f200d\_brt\_uds\_154 & Low-z (T2)/c & 4.56 (\Romannum{2}) & $2.61_{-0.20}^{+0.22}$ & $5.86_{-0.17}^{+0.17}$ & $2.83_{-1.13}^{+1.13}$ & $32_{-18}^{+56}$/$915\pm136$ & 1.0/$0.40\pm0.07$ & $9.23_{-0.37}^{+0.56}$/$10.37\pm0.09$ \\ 
       \hline
        f115d\_brt\_ceers\_146 & Low-z (T1)/d & N/A (\Romannum{3}) & $1.80_{-0.15}^{+0.15}$ & $2.64_{-0.06}^{+0.06}$ & $1.75_{-0.33}^{+0.33}$ & - & - & - \\ 
        \hline
    \end{tabular}
    }
    \raggedright
    \tablecomments{The Roman numerals in the parenthesis in the $z_{\rm spec}$ 
    column indicate the reliability of the spectroscopic redshifts: 
    \Romannum{1} -- $z_{\rm spec}$ determined using $\geq2$ high S/N emission 
    lines; 
    \Romannum{2} -- only one high S/N emission line is present, and 
    $z_{\rm spec}$ is determined by assigning the most probable line considering
    its $z_{\rm phot}$;  
    \Romannum{3} -- no emission line, i.e., no solid redshift can be 
    determined.  
    }
    \label{tab:spec-z}
\end{table*}

\section{Discussion}

\subsection{Mixture of EROs and High-$z$ Objects}

\begin{figure}
    \centering
    \includegraphics[width=0.95\linewidth]{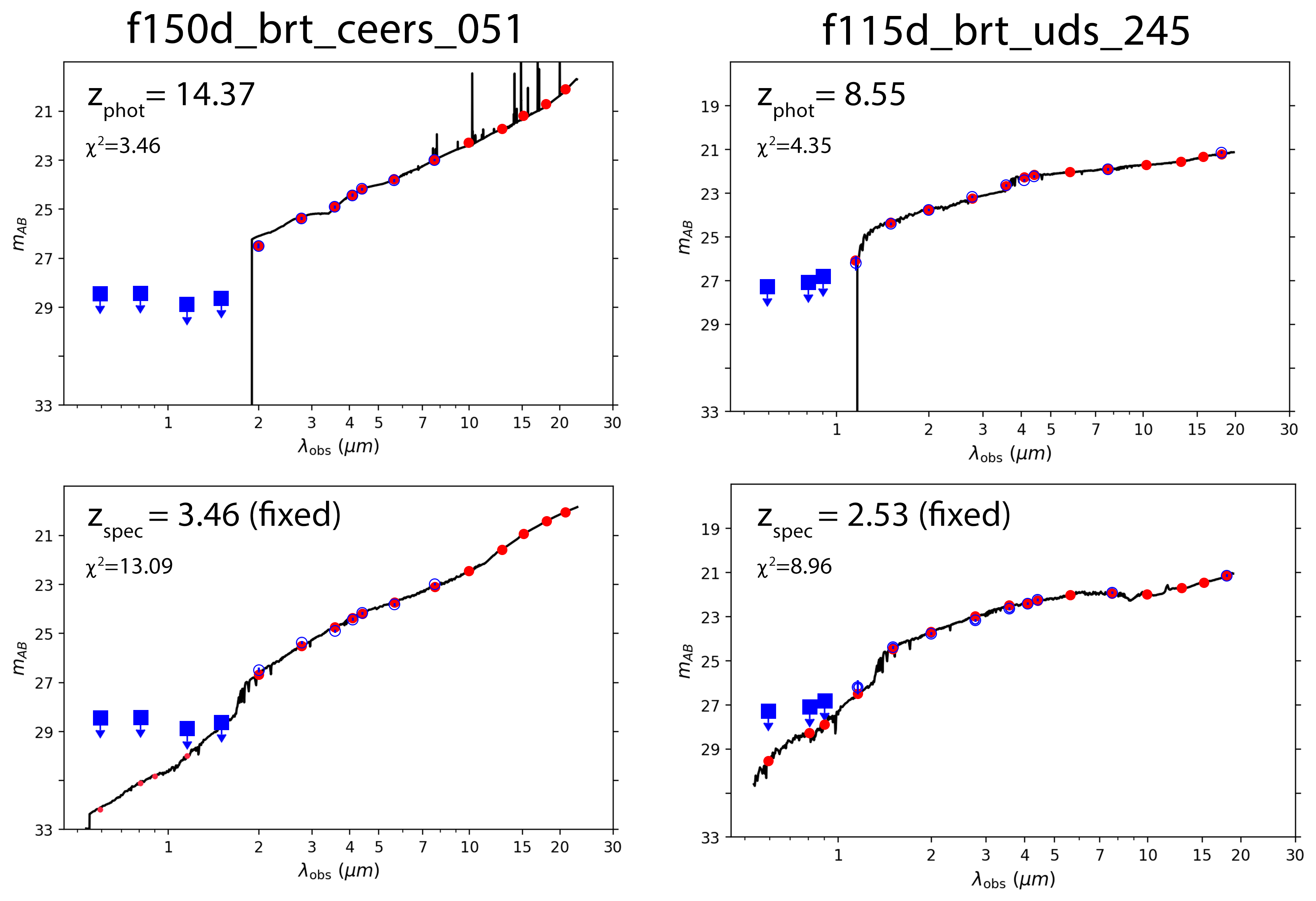}
    \caption{{Comparison of SED fittings (using CIGALE for demonstration) 
    for the two ``High-$z$'' candidates that were later confirmed as low 
    redshift ones. In both cases, the prominent Balmer break at low-$z$ mimics 
    the high-$z$ Lyman break signature very well. Nonetheless, the high-$z$ 
    solutions provide significantly better fits (as indicated by $\chi^2$),
    which demonstrates the difficulty in SED fitting.}
    }
    \label{fig:zsp_zph_compare}
\end{figure}

    Our SED analysis (see Section 5) shows that the bright dropouts through the
NIRCam bands should be dominated by red galaxies at low redshifts, and the limited
spectroscopic identifications (see Section 6) are consistent with this
interpretation. On the other hand, the SED analysis also shows that these bright
dropouts could contain a non-negligible fraction of real high-$z$ objects, which
is also confirmed by the spectroscopic identifications. The caveat is that the
SED screening for categorization is not ideal for these bright objects: 
our ``High-$z$'' category is severely contaminated, and there are still a large
fraction of objects that have to be placed in the ``Undecided'' category. 
We believe that this is just a manifest of the limitation of SED fitting as a method
in general: some low-$z$ galaxies can indeed have SEDs very similar to those of 
high-$z$ galaxies even when the photometry is extended to mid-IR. 
{To further demonstrate this limitation reflected in the two refuted
``High-$z$'' candidates, 
we ran CIGALE SED fitting using the 
same configuration as in Section~\ref{sec:sedtools} but with the redshift fixed at their $z_{\rm spec}$.
Figure~\ref{fig:zsp_zph_compare} compares their best-fit SED models when redshift is freed and fixed. 
Obviously, it is the strong Balmer break effectively mimicking 
the Lyman break signature that leads to the high-$z$ solution, which is a situation 
often encountered when the veto bands are not deep enough. In fact, the
high-$z$ solution gives a better fit in both cases, as judged by the $\chi^2$
values.
This is also seen in the two refuted sources 
in \citet[][]{Zavala23} and \citet[][]{AH23}: 
both appear in our supplement sample as {\tt f200d\_brt\_ceers\_264} and {\tt f150d\_brt\_ceers\_191}. 
Although they lie outside the MIRI footprint, their dropout signature is also caused by the strong Balmer break. 
}

{In short, it} is difficult for any SED fitting tools/templates to break 
the degeneracy {when lacking very deep veto band images. On the other 
hand, the high fraction of low-$z$ interlopers at the very bright end suggests
that apparent brightness itself could be used as a prior to pre-screen the
dropouts to retain a relatively pure high-$z$ sample, e.g., using 
$m_{\rm 356}\gtrsim25$--$26$~mag ($M_{\rm UV}\gtrsim-22$~mag), which has already 
been adopted in several selections
\citep[e.g., ][etc.]{Yan_highz_23,Adams24,Bouwens2023}.}


\subsection{Implication of bright dropouts at high-$z$ }

\begin{figure*}
    \centering
    \includegraphics[width=0.75\linewidth]{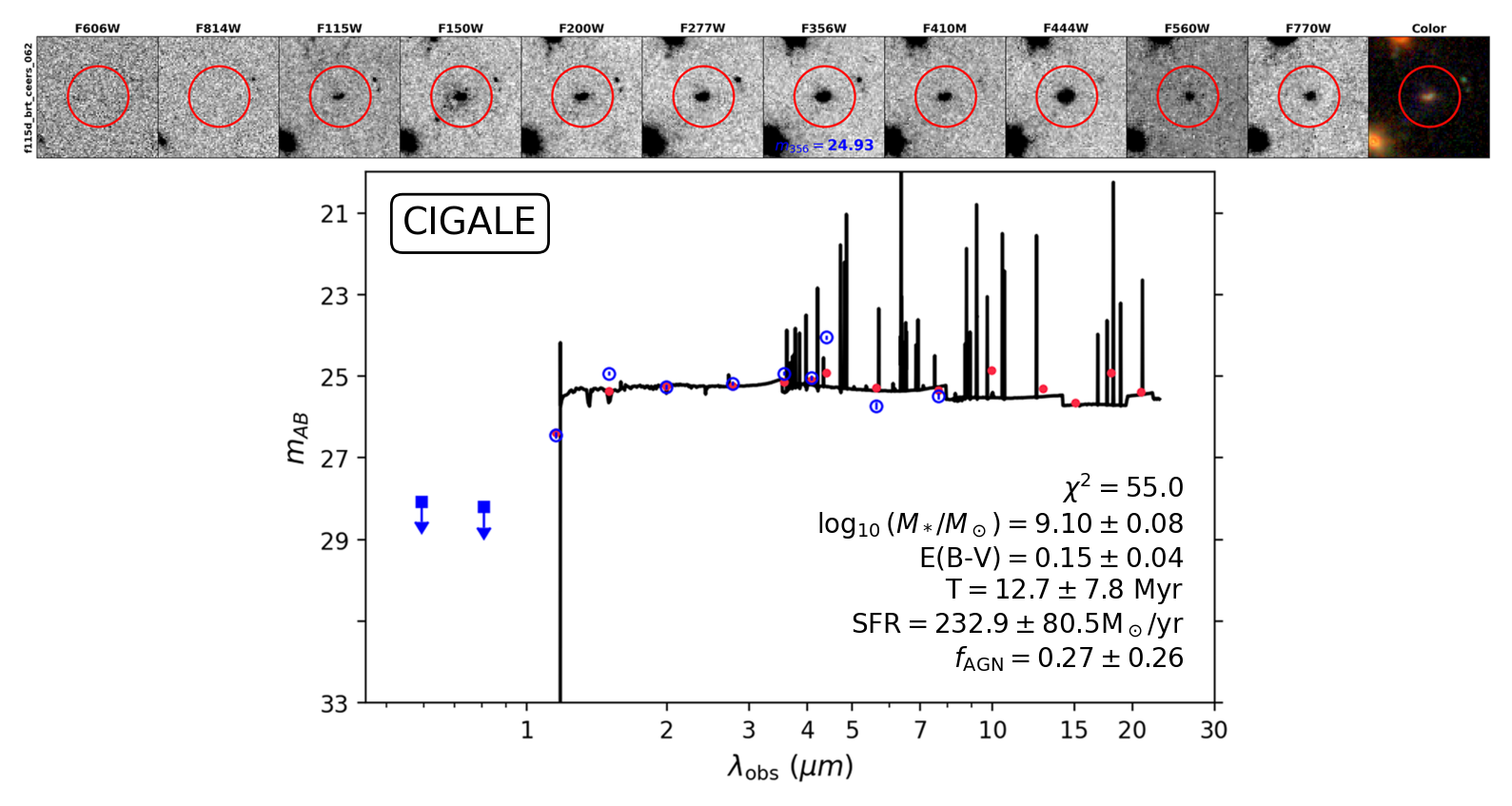}
    \caption{Image stamps of \texttt{f115d\_brt\_ceers\_062} (top) and its SED
    fitting results from CIGALE (bottom). This is a known galaxy at 
    $z_{\rm spec}=8.679$ recovered in our selection. The image stamps are similar 
    to those shown in Figure~\ref{fig:dropouts}. The SED fitting plot is similar 
    to those shown in Figure~\ref{fig:sed_example}, but is obtained by running
    CIGALE at the fixed $z=8.679$. The most important output physical parameters
    are shown.}
    \label{fig:f115d_ceers_062_sed}
\end{figure*}

   Bearing the aforementioned caveat in mind, we briefly discuss the implications 
of the possibility that some of these very bright dropouts could indeed be at 
high-$z$. If they are at $z>6$ as their $z_{\rm phot}$ indicate, they must be the 
most extraordinary galaxies in the early universe. The recovered high-$z$ object, 
\texttt{f115d\_brt\_ceers\_062} at $z=8.679$
\citep[see also][for the previous JWST identifications of the same object]{Tang2023, Larson2023a, Isobe2023, AH2023a},
is a good example. Interestingly, it has $m_{115}-m_{356}=1.52$~mag, which does
not meet the ERO criterion that we adopt here but is very close.
This galaxy was noted for its unusual brightness at such a 
high redshift and even more so for its very red color between the HST WFC3 F160W 
band ($\sim$1.6~$\mu$m) and the Spitzer IRAC Channel 2 ($\sim$4.5~$\mu$m), 
the latter of which was attributed to the strong [O~\Romannum{3}] emission lines 
being shifted to $\sim$4.8~$\mu$m \citep[][]{Zitrin2015b}. The JWST spectrum 
confirms that it is indeed the case. This indicates that the galaxy has very
active ongoing star formation, which is the reason that it is so luminous in the
rest frame UV ($M(1500{\rm \AA})=-22.4$~mag). As shown in
Figure~\ref{fig:f115d_ceers_062_sed}, the SED fitting by CIGALE at its 
$z_{\rm spec}$ gives the instantaneous SFR~$=232.9\pm80.5~M_\odot~{\rm yr}^{-1}$, 
which makes it qualified as a starburst. This extremely high SFR is the result of 
its high stellar mass ($M_*=10^{9.1}M_\odot$) and very young age 
($T=12.7\pm7.8$~Myr), the latter of which is needed to explain its very blue UV 
emission. In fact, even with such an extreme solution, its $m_{150}-m_{200}$ 
color is still not well explained; it is likely that an initial mass function 
more top-heavy than currently adopted will be necessary 
{\citep[e.g.,][]{Mason2023}}.

{Recent theoretical models could explain the number density of high-$z$ UV-
luminous galaxies to some extent \citep[see e.g.,][]{Dekel2023, Ferrara2023,
Shen2023}. However, it is still challenging for these new models to produce a
number density at $M_{\rm UV}\lesssim -23$~mag that could lead to positive 
detections within the volume that has been surveyed by the JWST. Our ``High-$z$'' 
categories contain a significant number of objects at $M_{\rm UV}< -23.5$~mag
if they are indeed at high redshifts. Of course, their high luminosities could
be explained if they host AGNs \citep[see e.g.,][]{Labbe2023}; however, one would
then need to explain the existence of supermassive black holes in such early
times.}


{Therefore, we argue that these extreme objects should not be simply ignored.
On the contrary, spectroscopic identification of them (confirmation or rejection)
will provide a valuable test on the early galaxy formation theories. If confirmed, 
}
the most challenging 
{aspect} will be explaining their stellar masses. Given their brightnesses in the two
reddest NIRCam bands (F356W and F444W), not only starburst-like SFRs but also
very high stellar masses will be required to fit their SEDs reasonably. Some of 
our objects will have to have $M_*\sim 10^{11}M_\odot$, which, to our knowledge, 
cannot be produced in {any new models} in the early universe. 

\section{Summary}

    We present a systematic study of very bright dropouts in the successive JWST 
NIRCam passbands, which was carried out using the public data in four blank 
fields over 500~arcmin$^2$. These objects were selected following the classic 
dropout method that is widely used to search for Lyman-break galaxies at high 
redshifts, with the only additional requirement that they must be very bright: 
the F090W and F115W dropouts must have $m_{356}\leq25.1$~mag, and the F150W and 
F200W dropouts must have $m_{356}\leq26.0$~mag. In total, 300 very 
bright dropouts were selected. We then focused on the 137 objects that fall within the 
coverage of the MIRI observations ($\sim$200~arcmin$^2$ of overlapping area), 
which form the main sample of this work. The rationale was that the inclusion of
the mid-IR measurements would constrain the SEDs more stringently. Using the 
fiducial criterion of $m_{115}-m_{356} > 2.0$~mag for the ERO selection, the 
vast majority of them (81\%) would qualify as EROs, although the usual ERO 
selection does not impose the criterion of non-detections in the veto bands as
the dropout selection does. The goal of this study is to understand the nature of 
such very bright dropouts, in particular, how they overlap with the ERO 
population at low redshifts and whether any of them could be high-$z$ LBGs as the 
dropout method means to select. 

   For this purpose, we used three different fitting tools to analyze their SEDs 
independently, each using a different set of fitting templates: Le Phare with the 
BC03 population synthesis models, EAZY with the spectra of 129 nearby galaxies 
from \cite{Brown14} that have mid-IR measurements, and CIGALE with the BC03 
models for stars plus the contributions from the nebular gas and the possible AGN 
component. Based on the derived $z_{\rm phot}$ values, we divided our objects into 
two categories, the ``High-$z$'' category containing objects with 
$z_{\rm phot}\geq 6.0$ and the ``Low-$z$'' category containing those
with $z_{\rm phot}<6.0$. In each category, we further sorted the objects into Tier 
1 and Tier 2 based on the goodness of fits and the internal consistency from three
SED fitting tools. Any objects that could not be classified as ``High-$z$'' or
``Low-$z$'' were put in the ``Undecided'' category. In the end, there are 10 
objects in ``High-$z$'' (80\% qualify as EROs), 93 in ``Low-$z$'' (87\% EROs) and 
34 in ``Undecided'' (63\% EROs). Therefore, our main conclusions are that (1) the 
NIRCam-selected very-bright dropouts are predominantly ($>67.9\%$) low-$z$ 
galaxies and (2) a non-negligible fraction ($>7\%$) of them could still be at 
high-$z$.

   Ten of our objects have existing JWST NIRSpec spectroscopic observations,
and we have obtained secure redshifts for seven of them. Among these seven
objects, six are at $z\approx 3$ (including two objects in our ``High-$z$'' 
category), and one is the recovery of a known galaxy at $z=8.679$ that is a 
Tier~2 object in our ``High-$z$'' category. These identifications, while still 
very limited, are consistent with our conclusions above. Although the
``High-$z$'' category is severely contaminated at such a bright level despite 
the inclusion of the mid-IR data, it does contain at least one genuine high-$z$ 
object. {Most objects in the ``High-$z$'' category are brighter than this 
one, and if more of them are confirmed to be at high-$z$, they would be in a 
very high luminosity regime still largely unexplored by the current theoretical 
models, and explaining their number density and extreme properties (e.g., very 
high stellar masses and SFRs) could pose a challenge.
}

\begin{acknowledgments}

  The authors thank the referee for the constructive comments that improve 
the quality of this paper. We acknowledge the support from the University of 
Missouri Research Council grant URC-23-029 and the NSF grant AST-2307447. 

{The JWST data presented in this article were obtained from the Mikulski Archive for Space Telescope (MAST) 
at the Space Telescope Science Institute. 
The specific observations analyzed can be accessed via \dataset[doi: 10.17909/8d3r-4p04]{https://doi.org/10.17909/8d3r-4p04}. 
}

\end{acknowledgments}

\bibliography{sample631}{}
\bibliographystyle{aasjournal}

\appendix
\counterwithin{figure}{section}
\counterwithin{table}{section}

\section{The Supplement Sample of Very Bright Dropouts}

Among the 300 very bright dropouts, 163 of them are outside of the MIRI coverage,
which we did not perform SED analysis due to the lack of the mid-IR data. These
objects form our supplement sample, which are listed below.

\startlongtable
\begin{longrotatetable}
\begin{deluxetable}{lccc||lccc||lccc}
\tablecolumns{12}
\tabletypesize{\tiny}
\tablecaption{Objects that do not have MIRI coverage. }
\tablehead{
\colhead{SID} & 
\colhead{R.A.} & 
\colhead{Decl.} & 
\colhead{$m_{356}$} & 
\colhead{SID} & 
\colhead{R.A.} & 
\colhead{Decl.} & 
\colhead{$m_{356}$} & 
\colhead{SID} & 
\colhead{R.A.} & 
\colhead{Decl.} & 
\colhead{$m_{356}$} 
}
\startdata
f115d\_brt\_ceers\_021 & 215.0885078 & 52.918795 & 23.80 & f115d\_brt\_ceers\_022 & 214.9938146 & 52.8521078 & 23.48 & f115d\_brt\_ceers\_031 & 215.0617498 & 52.9009611 & 21.76 \\
f115d\_brt\_ceers\_050 & 214.9806545 & 52.848635 & 22.84 & f115d\_brt\_ceers\_056 & 214.8544341 & 52.7595908 & 22.82 & f115d\_brt\_ceers\_059 & 215.0804619 & 52.9215632 & 24.15 \\
f115d\_brt\_ceers\_072 & 214.9396063 & 52.8255474 & 24.90 & f115d\_brt\_ceers\_074 & 215.0731942 & 52.9214419 & 24.04 & f115d\_brt\_ceers\_096 & 214.8097457 & 52.7397 & 23.39 \\
f115d\_brt\_ceers\_098 & 214.9438718 & 52.8358337 & 24.70 & f115d\_brt\_ceers\_148 & 214.9047696 & 52.8172013 & 22.62 & f115d\_brt\_ceers\_186 & 214.8607121 & 52.7968492 & 22.97 \\
f115d\_brt\_ceers\_189 & 214.8125544 & 52.7627987 & 24.52 & f115d\_brt\_ceers\_196 & 214.8852216 & 52.8157635 & 22.98 & f115d\_brt\_ceers\_277 & 214.8098893 & 52.7894358 & 24.05 \\
f115d\_brt\_ceers\_305 & 214.886832 & 52.8553988 & 24.96 & f115d\_brt\_ceers\_322 & 214.8099731 & 52.8097647 & 23.03 & f115d\_brt\_ceers\_327 & 214.9476307 & 52.911126 & 24.63 \\
f115d\_brt\_ceers\_328 & 214.876975 & 52.8604127 & 22.76 & f115d\_brt\_ceers\_360 & 214.7999771 & 52.8221046 & 24.08 & f115d\_brt\_ceers\_367 & 215.0368278 & 52.9935023 & 23.76 \\
f115d\_brt\_ceers\_371 & 214.9774813 & 52.9534984 & 23.16 & f115d\_brt\_ceers\_374 & 214.8505946 & 52.8660419 & 22.99 & f115d\_brt\_ceers\_392 & 214.7914996 & 52.8380477 & 22.07 \\
f115d\_brt\_ceers\_397 & 214.8357213 & 52.8753275 & 22.95 & f115d\_brt\_ceers\_408 & 214.8807006 & 52.9129751 & 23.27 & f115d\_brt\_ceers\_414 & 214.9040497 & 52.9327088 & 22.72 \\
f115d\_brt\_ceers\_422 & 214.8145038 & 52.8701449 & 24.80 & f115d\_brt\_ceers\_424 & 214.8381827 & 52.8888725 & 22.91 & f115d\_brt\_ceers\_453 & 214.8521098 & 52.9097693 & 22.87 \\
f115d\_brt\_ceers\_456 & 214.7634114 & 52.8478132 & 23.42 & f150d\_brt\_ceers\_036 & 214.8906508 & 52.8030673 & 24.57 & f150d\_brt\_ceers\_039 & 214.8829258 & 52.798169 & 24.32 \\
f150d\_brt\_ceers\_104 & 214.723005 & 52.7397616 & 24.28 & f150d\_brt\_ceers\_118 & 214.7180938 & 52.7480989 & 24.40 & f150d\_brt\_ceers\_148 & 214.9440605 & 52.9297443 & 26.15 \\
f150d\_brt\_ceers\_191 & 214.9145578 & 52.9430309 & 26.78 & f150d\_brt\_ceers\_201 & 214.9161447 & 52.9519038 & 24.85 & f150d\_brt\_ceers\_046 & 215.0505106 & 52.9260662 & 24.44 \\
f200d\_brt\_ceers\_222 & 214.9315925 & 52.9210116 & 24.80 & f200d\_brt\_ceers\_264 & 214.9091331 & 52.9372118 & 25.93 & f090d\_brt\_cosmos\_014 & 150.1440391 & 2.1772542 & 24.67 \\
f090d\_brt\_cosmos\_018 & 150.1138723 & 2.1819308 & 21.84 & f090d\_brt\_cosmos\_044 & 150.1449324 & 2.1912166 & 25.05 & f090d\_brt\_cosmos\_058 & 150.1431031 & 2.2004338 & 21.45 \\
f090d\_brt\_cosmos\_090 & 150.1217583 & 2.2130011 & 24.80 & f090d\_brt\_cosmos\_100 & 150.1244376 & 2.2172783 & 24.98 & f090d\_brt\_cosmos\_109 & 150.1236108 & 2.220721 & 23.89 \\
f090d\_brt\_cosmos\_137 & 150.1458027 & 2.2336448 & 23.46 & f090d\_brt\_cosmos\_143 & 150.1430039 & 2.2348342 & 23.69 & f090d\_brt\_cosmos\_198 & 150.1300109 & 2.2526974 & 21.99 \\
f090d\_brt\_cosmos\_199 & 150.1880081 & 2.2533098 & 23.56 & f090d\_brt\_cosmos\_209 & 150.1802681 & 2.2559495 & 23.51 & f090d\_brt\_cosmos\_252 & 150.1257594 & 2.2665978 & 24.13 \\
f090d\_brt\_cosmos\_284 & 150.089659 & 2.2758037 & 22.15 & f090d\_brt\_cosmos\_329 & 150.1393874 & 2.2822976 & 23.13 & f090d\_brt\_cosmos\_338 & 150.1859316 & 2.2831453 & 21.29 \\
f090d\_brt\_cosmos\_361 & 150.1816122 & 2.2895056 & 23.48 & f090d\_brt\_cosmos\_392 & 150.1394278 & 2.296584 & 23.96 & f090d\_brt\_cosmos\_421 & 150.1477821 & 2.3007552 & 22.50 \\
f090d\_brt\_cosmos\_454 & 150.1706349 & 2.307604 & 21.96 & f090d\_brt\_cosmos\_457 & 150.0759893 & 2.3096663 & 24.27 & f090d\_brt\_cosmos\_476 & 150.1121534 & 2.3140194 & 21.37 \\
f090d\_brt\_cosmos\_477 & 150.1132457 & 2.3149776 & 24.47 & f090d\_brt\_cosmos\_479 & 150.1307254 & 2.3141032 & 21.68 & f090d\_brt\_cosmos\_485 & 150.1292412 & 2.3163196 & 22.16 \\
f090d\_brt\_cosmos\_593 & 150.0990328 & 2.3436285 & 24.96 & f090d\_brt\_cosmos\_628 & 150.078639 & 2.3523408 & 24.40 & f090d\_brt\_cosmos\_633 & 150.0776435 & 2.3530061 & 24.58 \\
f090d\_brt\_cosmos\_690 & 150.0712848 & 2.3605719 & 23.72 & f090d\_brt\_cosmos\_710 & 150.072502 & 2.363681 & 24.74 & f090d\_brt\_cosmos\_721 & 150.0985811 & 2.3653675 & 21.09 \\
f090d\_brt\_cosmos\_737 & 150.1293022 & 2.3695651 & 21.33 & f090d\_brt\_cosmos\_749 & 150.1102552 & 2.3741219 & 21.57 & f090d\_brt\_cosmos\_753 & 150.0696261 & 2.3750253 & 23.42 \\
f090d\_brt\_cosmos\_767 & 150.085375 & 2.3801667 & 22.78 & f090d\_brt\_cosmos\_772 & 150.0698461 & 2.3815248 & 22.68 & f090d\_brt\_cosmos\_843 & 150.0823792 & 2.4033897 & 21.42 \\
f090d\_brt\_cosmos\_873 & 150.0952557 & 2.4233551 & 22.49 & f090d\_brt\_cosmos\_909 & 150.1307867 & 2.4443966 & 24.52 & f090d\_brt\_cosmos\_937 & 150.1269678 & 2.4653668 & 21.32 \\
f115d\_brt\_cosmos\_025 & 150.1197606 & 2.1886 & 23.63 & f115d\_brt\_cosmos\_107 & 150.1536777 & 2.2477695 & 23.00 & f115d\_brt\_cosmos\_119 & 150.1092878 & 2.252688 & 23.39 \\
f115d\_brt\_cosmos\_162 & 150.1129588 & 2.2745327 & 23.79 & f115d\_brt\_cosmos\_335 & 150.096574 & 2.3520867 & 24.90 & f115d\_brt\_cosmos\_425 & 150.0864335 & 2.3953681 & 24.28 \\
f150d\_brt\_cosmos\_004 & 150.1289676 & 2.1741546 & 24.40 & f150d\_brt\_cosmos\_088 & 150.1365424 & 2.2605942 & 23.29 & f150d\_brt\_cosmos\_094 & 150.1510636 & 2.2623562 & 22.43 \\
f150d\_brt\_cosmos\_320 & 150.0935145 & 2.3947138 & 24.44 & f150d\_brt\_cosmos\_330 & 150.0925514 & 2.3981518 & 24.25 & f150d\_brt\_cosmos\_338 & 150.11263 & 2.4066739 & 23.88 \\
f150d\_brt\_cosmos\_352 & 150.1206107 & 2.4180919 & 23.77 & f150d\_brt\_cosmos\_362 & 150.1165873 & 2.4248924 & 23.98 & f150d\_brt\_cosmos\_371 & 150.1057684 & 2.4348203 & 24.23 \\
f090d\_brt\_jsdeep\_502 & 53.1549052 & -27.7307764 & 22.86 & f090d\_brt\_jsmed\_132 & 53.14957 & -27.8769106 & 23.19 & f090d\_brt\_jsmed\_181 & 53.1466098 & -27.871024 & 23.20 \\
f115d\_brt\_jsmed\_262 & 53.0410562 & -27.8377208 & 23.82 & f150d\_brt\_jsmed\_001 & 53.0413528 & -28.0261991 & 22.93 & f150d\_brt\_jsmed\_003 & 53.0377722 & -28.0226959 & 23.71 \\
f150d\_brt\_jsmed\_004 & 53.0445866 & -28.0217799 & 23.68 & f150d\_brt\_jsmed\_009 & 53.0814964 & -28.0060328 & 23.06 & f150d\_brt\_jsmed\_020 & 53.0568695 & -27.9719865 & 24.66 \\
f150d\_brt\_jsmed\_242 & 53.1026077 & -27.806501 & 24.87 & f200d\_brt\_jsmed\_004 & 53.066538 & -28.0059346 & 25.84 & f090d\_brt\_uds\_021 & 34.3969829 & -5.282103 & 23.75 \\
f090d\_brt\_uds\_130 & 34.3220264 & -5.2457358 & 21.28 & f090d\_brt\_uds\_139 & 34.4151013 & -5.2365206 & 22.53 & f090d\_brt\_uds\_144 & 34.4979806 & -5.2293821 & 22.56 \\
f090d\_brt\_uds\_159 & 34.3708622 & -5.2097999 & 22.55 & f090d\_brt\_uds\_216 & 34.2291325 & -5.1644344 & 23.48 & f115d\_brt\_uds\_004 & 34.340531 & -5.320432 & 22.29 \\
f115d\_brt\_uds\_009 & 34.5132061 & -5.3191778 & 23.38 & f115d\_brt\_uds\_012 & 34.2461584 & -5.3183936 & 24.10 & f115d\_brt\_uds\_039 & 34.515515 & -5.3103855 & 23.92 \\
f115d\_brt\_uds\_048 & 34.3444628 & -5.3050216 & 23.64 & f115d\_brt\_uds\_055 & 34.4625563 & -5.3040264 & 23.80 & f115d\_brt\_uds\_059 & 34.5231533 & -5.3030314 & 23.80 \\
f115d\_brt\_uds\_065 & 34.255683 & -5.3017361 & 24.02 & f115d\_brt\_uds\_213 & 34.3018505 & -5.2646227 & 23.94 & f115d\_brt\_uds\_324 & 34.3876199 & -5.2383642 & 22.79 \\
f115d\_brt\_uds\_335 & 34.3040498 & -5.2366875 & 21.16 & f115d\_brt\_uds\_346 & 34.2558777 & -5.2338276 & 21.77 & f115d\_brt\_uds\_349 & 34.4951591 & -5.2323406 & 23.22 \\
f115d\_brt\_uds\_357 & 34.2313445 & -5.2290864 & 24.35 & f115d\_brt\_uds\_364 & 34.2692778 & -5.2280467 & 22.29 & f115d\_brt\_uds\_416 & 34.2959616 & -5.2127592 & 23.82 \\
f115d\_brt\_uds\_436 & 34.2409063 & -5.2055225 & 23.49 & f115d\_brt\_uds\_451 & 34.3081378 & -5.2035669 & 24.34 & f115d\_brt\_uds\_465 & 34.40423 & -5.2008092 & 22.55 \\
f115d\_brt\_uds\_495 & 34.2351048 & -5.1944463 & 22.96 & f115d\_brt\_uds\_500 & 34.3716914 & -5.19226 & 22.81 & f115d\_brt\_uds\_537 & 34.4834163 & -5.1786782 & 21.89 \\
f115d\_brt\_uds\_582 & 34.3806723 & -5.1650075 & 22.39 & f115d\_brt\_uds\_593 & 34.3436309 & -5.1598316 & 22.94 & f115d\_brt\_uds\_794 & 34.2278043 & -5.0905055 & 23.58 \\
f115d\_brt\_uds\_797 & 34.3886645 & -5.0868892 & 24.93 & f115d\_brt\_uds\_314 & 34.2432658 & -5.239754 & 23.97 & f115d\_brt\_uds\_350 & 34.3230504 & -5.2314974 & 24.73 \\
f115d\_brt\_uds\_816 & 34.4150182 & -5.236605 & 25.09 & f150d\_brt\_uds\_010 & 34.3512775 & -5.3176514 & 23.91 & f150d\_brt\_uds\_147 & 34.3383058 & -5.2539202 & 23.95 \\
f150d\_brt\_uds\_195 & 34.2483377 & -5.237603 & 24.18 & f150d\_brt\_uds\_196 & 34.4152814 & -5.2363373 & 24.74 & f150d\_brt\_uds\_197 & 34.2596709 & -5.2345851 & 25.11 \\
f150d\_brt\_uds\_207 & 34.3001351 & -5.2301039 & 23.51 & f150d\_brt\_uds\_236 & 34.279762 & -5.2150263 & 24.48 & f150d\_brt\_uds\_264 & 34.4928449 & -5.1957711 & 24.60 \\
f150d\_brt\_uds\_282 & 34.325191 & -5.184433 & 24.18 & f150d\_brt\_uds\_289 & 34.5235392 & -5.1804038 & 24.75 & f150d\_brt\_uds\_291 & 34.4908477 & -5.1794282 & 24.66 \\
f150d\_brt\_uds\_389 & 34.2716179 & -5.1282498 & 25.68 & f150d\_brt\_uds\_391 & 34.2714136 & -5.1271676 & 25.44 & f150d\_brt\_uds\_443 & 34.3813041 & -5.0919038 & 24.12 \\
f150d\_brt\_uds\_447 & 34.4178077 & -5.0894718 & 24.82 & f150d\_brt\_uds\_452 & 34.3945399 & -5.0776618 & 22.70 & f150d\_brt\_uds\_187 & 34.2432052 & -5.2397594 & 25.54 \\
f150d\_brt\_uds\_279 & 34.4877694 & -5.1845142 & 24.52 & f200d\_brt\_uds\_087 & 34.4940772 & -5.2783977 & 25.20 & f200d\_brt\_uds\_183 & 34.3763787 & -5.2370853 & 25.67 \\
f200d\_brt\_uds\_518 & 34.2105 & -5.0931469 & 25.61 &  &  &  &  &  &  &  &  \\
\hline
\enddata
\end{deluxetable}
\label{tab:suppl}
\end{longrotatetable}

\section{Very Bright Dropouts in the ``Undecided'' Category in the Main Sample}

   The objects in the main sample that cannot be placed in either the
``High-$z$'' or the ``Low-$z$'' categories were assigned as ``Undecided'', which
are given in the table below.

\begin{table*}[hbt!]
\centering
\scriptsize
\begin{tabular}{lccc||lccc}
\hline
SID & R.A. & Decl. & $m_{356}$ & SID & R.A. & Decl. & $m_{356}$ \\
\hline
f115d\_brt\_ceers\_430 & 214.751589 & 52.8299477 & 24.90 & f150d\_brt\_ceers\_113 & 214.8296653 & 52.8207925 & 24.11 \\
f150d\_brt\_ceers\_163 & 214.7678858 & 52.8163009 & 25.70 & f090d\_brt\_cosmos\_244 & 150.0775622 & 2.264394 & 25.13 \\
f090d\_brt\_cosmos\_351 & 150.0850761 & 2.2854155 & 23.15 & f090d\_brt\_cosmos\_363 & 150.0708811 & 2.2893357 & 23.40 \\
f090d\_brt\_cosmos\_395 & 150.0994053 & 2.2972383 & 23.20 & f090d\_brt\_cosmos\_419 & 150.098576 & 2.3012176 & 23.10 \\
f090d\_brt\_cosmos\_498 & 150.1762613 & 2.3194308 & 24.27 & f090d\_brt\_cosmos\_790 & 150.1835939 & 2.3903725 & 25.12 \\
f090d\_brt\_cosmos\_796 & 150.1557834 & 2.392438 & 24.52 & f090d\_brt\_cosmos\_919 & 150.1642394 & 2.4531855 & 23.51 \\
f090d\_brt\_cosmos\_927 & 150.1848911 & 2.4604111 & 25.25 & f115d\_brt\_cosmos\_010 & 150.0971618 & 2.1745523 & 23.54 \\
f115d\_brt\_cosmos\_344 & 150.14326 & 2.3560209 & 23.04 & f150d\_brt\_cosmos\_231 & 150.1819653 & 2.3355567 & 23.68 \\
f150d\_brt\_cosmos\_368 & 150.13919 & 2.4319799 & 21.61 & f150d\_brt\_cosmos\_394 & 150.1472663 & 2.4740686 & 24.45 \\
f090d\_brt\_jsdeep\_283 & 53.2063219 & -27.7757229 & 24.68 & f090d\_brt\_jsmed\_040 & 53.0811514 & -27.902613 & 25.09 \\
f090d\_brt\_jsmed\_126 & 53.0528574 & -27.8777309 & 23.45 & f090d\_brt\_jsmed\_174 & 53.0796299 & -27.870759 & 24.88 \\
f090d\_brt\_jsmed\_186 & 53.1015139 & -27.8699644 & 23.45 & f090d\_brt\_uds\_076 & 34.4887416 & -5.2656961 & 25.09 \\
f090d\_brt\_uds\_225 & 34.3544488 & -5.1593633 & 24.70 & f090d\_brt\_uds\_254 & 34.4897227 & -5.1457693 & 25.09 \\
f090d\_brt\_uds\_260 & 34.3822548 & -5.1435204 & 24.43 & f115d\_brt\_uds\_640 & 34.3650612 & -5.1488379 & 22.92 \\
f115d\_brt\_uds\_749 & 34.3749038 & -5.1129002 & 20.86 & f150d\_brt\_uds\_057 & 34.4969958 & -5.2899889 & 25.46 \\
f150d\_brt\_uds\_097 & 34.3248951 & -5.2712995 & 24.66 & f150d\_brt\_uds\_361 & 34.2744535 & -5.1437765 & 23.47 \\
f200d\_brt\_uds\_044 & 34.3411055 & -5.2964632 & 25.89 & f200d\_brt\_uds\_420 & 34.4054154 & -5.1393098 & 25.55 \\
\hline
\end{tabular}
\caption{Objects in the ``Undecided'' category.}
\label{tab:undecided}
\end{table*}

\end{document}